\documentclass[preprint,showpacs,amsmath,amssymb,aps,prd,nofootinbib]{revtex4}
\usepackage{epsfig,graphicx,color,appendix}

\begin{document}
\begin{flushright}
KIAS-P12045
\end{flushright}

\title{\mbox{}\\[10pt]
Simple renormalizable flavor symmetry for neutrino oscillations}

\author{Y. H. Ahn\footnote{Email: yhahn@kias.re.kr},
Seungwon Baek\footnote{Email: swbaek@kias.re.kr},
Paolo Gondolo\footnote{Email: paolo.gondolo@utah.edu. On sabbatical leave from University of Utah.}
}

\affiliation{School of Physics, KIAS, Seoul 130-722, Korea}

\date{\today}% It is always \today, today,

\begin{abstract}
The recent measurement of a non-zero neutrino mixing angle
$\theta_{13}$ requires a modification of
the tri-bimaximal mixing pattern that predicts a zero value for it. We propose a new neutrino mixing pattern based on a spontaneously-broken $A_{4}$ flavor symmetry and a type-I seesaw mechanism. Our model allows for approximate tri-bimaximal mixing and non-zero $\theta_{13}$, and contains a natural way to implement low and high energy CP violation in neutrino oscillations, and leptogenesis with a renormalizable Lagrangian.
Both normal and inverted mass hierarchies are permitted within $3\sigma$ experimental bounds, with the prediction of small (large) deviations from maximality in the atmospheric mixing angle for the normal (inverted) case. Interestingly, we show that the inverted case is excluded by the global analysis in $1\sigma$ experimental bounds, while the most recent MINOS data seem to favor the inverted case. Our model make predictions for the Dirac CP phase in the normal and inverted hierarchies,
which can be tested in near-future neutrino oscillation experiments.
Our model  also predicts  the effective mass $|m_{ee}|$ measurable in neutrinoless double beta decay to be in the range
$0.04\lesssim |m_{ee}| \lesssim 0.15$ eV for the normal hierarchy and $0.06\lesssim |m_{ee}| \lesssim 0.11$ eV for the inverted hierarchy,
both of which are within the sensitivity of the next generation experiments.
\end{abstract}

\maketitle %
%%%%%%%%%%%%%%%%%%%%%%%%%%%%%%%%%%%%%%%%%%%%%%%%%%%%%%%%%%%%%%%%%%%%%%%%%%%%%%%%%%%%%%%%%%
\section{Introduction}
The large values of the solar ($\theta_{12}\simeq35^\circ$) and atmospheric ($\theta_{23}\simeq45^\circ)$~\cite{PDG} neutrino mixing angles
 may be telling us about  new symmetries in the lepton sector not present in the quark sector, and may provide us with a clue to the nature of the quark-lepton physics beyond the standard model. Theoretically, a great deal of effort has been put into constructing flavor models with high predictive power, especially those giving the tri-bimaximal (TBM) mixing angles~\cite{TBM}:
 \begin{eqnarray}
\theta_{13}&=&0,\qquad\theta_{23}=\frac{\pi}{4}=45^\circ~,\qquad\theta_{12}=\sin^{-1}\left(\frac{1}{\sqrt{3}}\right)\simeq35.3^\circ~.
 \label{TBM}
 \end{eqnarray}
However, the Daya Bay and RENO collaborations~\cite{An:2012eh,Ahn:2012nd} have reported the first measurements of a non-zero value for the mixing angle $\theta_{13}$:
 \begin{eqnarray}
  \sin^2 2\theta_{13}=0.092\pm 0.016\text{(stat)}\pm 0.005\text{(syst)},
 \end{eqnarray}
 and
 \begin{eqnarray}
  \sin^2 2\theta_{13}=0.113\pm 0.013\text{(stat)}\pm 0.019\text{(syst)},
 \end{eqnarray}
respectively, corresponding to an angle $\theta_{13}\approx 9^\circ$. These results are in good agreement with the previous data from the T2K, MINOS and Double Chooz collaborations~\cite{Data}. A non-zero value of $\theta_{13}$ indicates that the TBM pattern for neutrino mixing should be modified. In addition, at the Neutrino 2012 conference in Kyoto, the MINOS Collaboration has announced a non-maximal value for the atmospheric mixing angle $\theta_{23}$~\cite{kyoto},
 \begin{eqnarray}
  \sin^2 2\theta_{23}=0.94^{+0.04}_{-0.05}\pm 0.04~,
 \label{minos}
 \end{eqnarray}
with maximal mixing disfavored at the $88\%$ C.L. This result, which  was not included the global analysis in~\cite{Tortola:2012te}, comes from the analysis of $\nu_{\mu}$ disappearance in the MINOS accelerator beam, and points to one of two possible values for $\theta_{23}$, namely $\theta_{23}=38^{\circ}$ or $\theta_{23}=52^{\circ}$. If it holds, this result also calls for a deviation from the tri-bimaximal mixing pattern.

Furthermore, the presence of  CP violation in the lepton sector is still unknown. Experimentally, CP violation may become observable in a future generation of neutrino oscillation experiments~[T2K, NO$\nu$A]~\cite{Nunokawa:2007qh}.  Theoretically,
a flavor symmetry that describes and explains the large reactor mixing angle $\theta_{13}\simeq9^{\circ}$ while keeping the TBM values $\theta_{23}\simeq45^{\circ}$ and $\theta_{12}\simeq35^{\circ}$ may originate in two ways:  (i) a large $\theta_{13}=\lambda_{C}/\sqrt{2}$, with $\lambda_{C}$  the Cabbibo angle, mainly governed by higher-order corrections in the charged lepton sector~\cite{Ahn:2011yj}, where the TBM pattern is a good zero-order approximation to reality, or (ii) a large $\theta_{13}$ from the neutrino sector itself through a new flavor symmetry without resorting to higher-order corrections in the charged lepton sector~\cite{Ahn:2012tv}.

In this paper, we propose a new and simple model for the lepton sector with $A_{4}$ flavor symmetry in the framework of a type-I seesaw mechanism.
It is different from  previous works using  $A_{4}$ flavor symmetries~\cite{Babu:2002dz, Altarelli:2005yp, A4, A4BO}\footnote{E.Ma and G.Rajasekaran~\cite{Ma:2001dn} have introduced for the first time the $A_{4}$ symmetry to avoid the mass degeneracy of $\mu$ and $\tau$ under a $\mu$--$\tau$ symmetry~\cite{mutau}.} in that the Dirac neutrino Yukawa coupling constants do not all have the same magnitude. Our model can naturally explain the TBM large value of $\theta_{13}$ and can also provide a possibility for low energy CP violation in neutrino oscillations with a renormalizable Lagrangian and small Yukawa coupling parameters, i.e.\ neutrino masses. The seesaw mechanism, besides explaining of smallness of the measured neutrino masses,  has another appealing feature: generating the observed baryon asymmetry in our Universe by means of  leptogenesis~\cite{review}. Since the conventional $A_{4}$ models realized with type-I or -III seesaw and a tree-level Lagrangian lead to an exact TBM and vanishing leptonic CP-asymmetries responsible for leptogenesis (due to the proportionality  of the $Y^{\dag}_{\nu}Y_{\nu}$ combination of the Dirac neutrino Yukawa matrix $Y_{\nu}$ to the unit matrix), authors usually introduce soft-breaking terms or higher-dimensional operators with many parameters, in order to explain the non-zero $\theta_{13}$ as well as the non-vanishing CP-asymmetries.

Our model is based on a renormalizable $SU(2)_L\times U(1)_Y\times A_{4}$ Lagrangian with minimal Yukawa couplings,
and gives rise to a non-degenerate Dirac neutrino Yukawa matrix and a unique CP-violation pattern.
This opens the possibility of explaining the non-zero value of $\theta_{13}\simeq9^{\circ}$ still
maintaining TBM  for the other two neutrino mixing angles $\theta_{23}\simeq45^{\circ}$ and
$\theta_{12}\simeq35^{\circ}$; furthermore, this allows an economic way to achieve low energy
CP violation in neutrino oscillations as well as high energy CP violation for leptogenesis.
%In addition, residual symmetries after the breaking of the $A_{4}$ flavor symmetry make
%the lightest particle charged under the $A_4$ symmetry stable, providing a dark matter candidate.

This paper is organized as follows. In the next section, we lay down the particle content
and the field representations under the $A_4$  flavor symmetry in our model, as well as explain the characteristic points of our model phenomenology at low and high energy. In Sec.~III, we present the neutrino mixing angles, and how the low energy CP violation could be generated in both normal and inverted mass hierarchies, including our predictions for neutrinoless double beta decay. We give our conclusions in Sec.~IV, and in Appendix A we outline the minimization of the scalar potential and the vacuum alignments.

%%%%%%%%%%%%%%%%%%%%%%%%%%%%%%%%%%%%%%%%%%%%%%%%%%%%%%%%%%%%%%%%%%%%%%%%%%%%%%%%%%%%%%%%%%
\section{flavor $A_{4}$ symmetry for non-zero $\theta_{13}$ and leptogenesis}

In the absence of flavor symmetries, particle masses and mixings
are generally undetermined in a gauge theory. Here, to understand the present non-zero $\theta_{13}$ and TBM angles ($\theta_{12}, \theta_{23}$) of the neutrino oscillation data and baryogenesis via leptogenesis, we propose a new discrete symmetry based on an $A_{4}$ flavor symmetry for leptons in a renormalizable Lagrangian.\footnote{To include the quark sector, the symmetry could
be promoted to the binary tetrahedral group $T'$~\cite{Case:1956zz}.}

The  group $A_{4}$ is the symmetry group of the
tetrahedron, isomorphic to the finite group of the even permutations of four
objects. The group $A_{4}$ has two generators, denoted $S$ and $T$, satisfying the relations $S^{2}=T^{3}=(ST)^{3}=1$. In the three-dimensional real representation, $S$ and $T$ are given by
 \begin{eqnarray}
 S={\left(\begin{array}{ccc}
 1 &  0 &  0 \\
 0 &  -1 & 0 \\
 0 &  0 &  -1
 \end{array}\right)}~,\qquad T={\left(\begin{array}{ccc}
 0 &  1 &  0 \\
 0 &  0 &  1 \\
 1 &  0 &  0
 \end{array}\right)}~.
 \label{generator}
 \end{eqnarray}
$A_{4}$ has four irreducible representations: one triplet ${\bf
3}$ and three singlets ${\bf 1}, {\bf 1}', {\bf 1}''$. An $A_4$ triplet $(a_1,a_2,a_3)$ transforms in the unitary representation by multiplication with the $S$ and $T$ matrices in Eq.~(\ref{generator}) above,
\begin{align}
S \begin{pmatrix} a_1 \\ a_2 \\ a_3 \end{pmatrix} = \begin{pmatrix} a_1 \\ -a_2 \\ -a_3 \end{pmatrix},
\qquad
T \begin{pmatrix} a_1 \\ a_2 \\ a_3 \end{pmatrix} = \begin{pmatrix} a_2 \\ a_3 \\ a_1 \end{pmatrix}.
\end{align}
An $A_4$ singlet $a$ is invariant under the action of $S$ ($Sa=a$), while the action of $T$ produces $Ta=a$ for ${\bf 1}$, $Ta=\omega a$ for ${\bf 1}'$, and $Ta=\omega^2 a$ for ${\bf 1}''$, where $\omega=e^{i2\pi/3}$ is a complex cubic-root of unity.
Products of two $A_4$ representations decompose into irreducible representations according to the following multiplication rules: ${\bf 3}\otimes{\bf 3}={\bf 3}_{s}\oplus{\bf
3}_{a}\oplus{\bf 1}\oplus{\bf 1}'\oplus{\bf 1}''$, ${\bf
1}'\otimes{\bf 1}''={\bf 1}$, ${\bf 1}'\otimes{\bf 1}'={\bf 1}''$
and ${\bf 1}''\otimes{\bf 1}''={\bf 1}'$. Explicitly, if $(a_{1},
a_{2}, a_{3})$ and $(b_{1}, b_{2}, b_{3})$ denote two $A_4$ triplets,
 \begin{eqnarray}
  (a\otimes b)_{{\bf 3}_{\rm s}} &=& (a_{2}b_{3}+a_{3}b_{2}, a_{3}b_{1}+a_{1}b_{3}, a_{1}b_{2}+a_{2}b_{1})~,\nonumber\\
  (a\otimes b)_{{\bf 3}_{\rm a}} &=& (a_{2}b_{3}-a_{3}b_{2}, a_{3}b_{1}-a_{1}b_{3}, a_{1}b_{2}-a_{2}b_{1})~,\nonumber\\
  (a\otimes b)_{{\bf 1}} &=& a_{1}b_{1}+a_{2}b_{2}+a_{3}b_{3}~,\nonumber\\
  (a\otimes b)_{{\bf 1}'} &=& a_{1}b_{1}+\omega a_{2}b_{2}+\omega^{2} a_{3}b_{3}~,\nonumber\\
  (a\otimes b)_{{\bf 1}''} &=& a_{1}b_{1}+\omega^{2} a_{2}b_{2}+\omega a_{3}b_{3}~.
 \end{eqnarray}

To make the presentation of our model physically more transparent, we define the $T$-flavor quantum number $T_f$ through the eigenvalues of the operator $T$, for which $T^3=1$. In detail, we say that a field $f$ has $T$-flavor $T_f=0$, +1, or -1 when it is an eigenfield of the $T$ operator with eigenvalue $1$, $\omega$, $\omega^2$, respectively (in short, with eigenvalue $\omega^{T_f}$ for $T$-flavor $T_f$, considering the cyclical properties of the cubic root of unity $\omega$). The $T$-flavor is an additive quantum number modulo 3. We also define the $S$-flavor-parity through the eigenvalues of the operator $S$, which are +1 and -1 since $S^2=1$, and we speak of $S$-flavor-even and $S$-flavor-odd fields.
For $A_4$-singlets, which are all $S$-flavor-even, the $\mathbf{1}$ representation has no $T$-flavor ($T_f=0$), the $\mathbf{1}'$ representation has $T$-flavor $T_f=+1$, and the $\mathbf{1}''$ representation has $T$-flavor $T_f=-1$. Since for $A_4$-triplets, the operators $S$ and $T$ do not commute, $A_4$-triplet fields cannot simultaneously have a definite $T$-flavor and a definite $S$-flavor-parity.
While the real representation of $A_4$ in Eqs.~(\ref{generator}), in which $S$ is diagonal, is useful in writing the Lagrangian, the physical meaning of our model is more apparent in the $T$-flavor representation in which $T$ is diagonal. This representation is obtained through the unitary transformation
\begin{align}
A \to A'=U_{\omega}AU^{\dag}_{\omega},
\end{align}
where $A$ is any $A_4$ matrix in the real representation and
\begin{align}
U_{\omega}=\frac{1}{\sqrt{3}}{\left(\begin{array}{ccc}
 1 &  1 &  1 \\
 1 & \omega^{2} & \omega \\
 1 & \omega & \omega^{2}
 \end{array}\right)}.
 \label{eq:Uomega}
\end{align}
We have
 \begin{eqnarray}
 S'=\frac{1}{3} \, {\left(\begin{array}{ccc}
 -1 &  2 &  2 \\
 2 &  -1 & 2 \\
 2 &  2 &  -1
 \end{array}\right)}~,\qquad T'={\left(\begin{array}{ccc}
 1 &  0 &  0 \\
 0 &  \omega &  0 \\
 0 &  0 &  \omega^2
 \end{array}\right)}~.
 \label{generator2}
 \end{eqnarray}
Despite the physical advantages of the $S'$, $T'$ representation, for clarity of exposition and to avoid confusion and complications, in this paper  we use the real representation $S$, $T$ almost exclusively. For reference,
%in the real representation, the $T$-flavor eigentriplets with $T$-flavor $T_f=0$, +1, and -1 are, respectively,
%\begin{align}
%a_0 = \frac{1}{\sqrt{3}} \begin{pmatrix} 1 \\ 1 \\ 1 \end{pmatrix} ,
%\quad
%a_{+1} = \frac{1}{\sqrt{3}} \begin{pmatrix} 1 \\ \omega \\ \omega^2 \end{pmatrix} ,
%\quad
%a_{-1}= \frac{1}{\sqrt{3}} \begin{pmatrix} 1 \\ \omega^2 \\ \omega \end{pmatrix} .
%\end{align}
%An
an $A_4$ triplet field with components $(a_1,a_2,a_3)$ in the real representation can be expressed in terms of $T$-flavor eigenfields $(a_e,a_\mu,a_\tau)$ (the notation comes from our lepton assignments below) as
\begin{align}
 a_{1} = \frac{a_{e}+a_{\mu}+a_{\tau}}{\sqrt{3}} , \quad
 a_{2} = \frac{a_{e}+\omega^2 a_{\mu}+\omega a_{\tau}}{\sqrt{3}} , \quad
 a_{3} = \frac{a_{e}+\omega a_{\mu}+\omega^2 a_{\tau}}{\sqrt{3}} .
 \label{eq:Ua1}
\end{align}
Inversely,
 \begin{align}
 a_{e}  = \frac{a_{1}+a_{2}+a_{3}}{\sqrt{3}} , \quad
 a_{\mu}  = \frac{a_{1}+\omega a_{2}+\omega^2 a_{3}}{\sqrt{3}} , \quad
 a_{\tau}  = \frac{a_{1}+\omega^2 a_{2}+\omega a_{3}}{\sqrt{3}} .
 \label{eq:Ua2}
 \end{align}

We extend the standard model (SM) by the inclusion of an $A_4$-triplet of
right-handed $SU(2)_{L}$-singlet Majorana neutrinos $N_{R}$, and the introduction of two types of scalar Higgs fields  besides the usual SM $SU(2)_{L}$-doublet Higgs bosons $\Phi$, which we take to be an $A_4$-singlet with no $T$-flavor ($\mathbf{1}$ representation): a second $SU(2)_{L}$-doublet of Higgs bosons $\eta$, which is distinguished from $\Phi$ by being an $A_4$-triplet, and an $SU(2)_{L}$-singlet $A_4$-triplet real scalar field $\chi$:
 \begin{eqnarray}
  \Phi =
\begin{pmatrix} \varphi^{+} \\ \varphi^{0} \end{pmatrix},
\qquad
\eta_j =
\begin{pmatrix}
  \eta^{+}_j \\
  \eta^{0}_j
\end{pmatrix} ,
\qquad
\chi_j,
\qquad
j=1,2,3.
  \label{Higgs}
 \end{eqnarray}

We assign each flavor of leptons to one of the three $A_4$ singlet representations: the electron-flavor to the ${\bf 1}$ ($T$-flavor 0), the muon flavor to the ${\bf 1}'$  ($T$-flavor +1), and the tau flavor to the ${\bf 1}''$  ($T$-flavor -1). (Note in this respect that our $A_4$ flavor group is not a symmetry under exchange of any two lepton flavors, like $e$ and $\mu$, for example. Our $A_4$ flavor group is implemented as a global symmetry of the Lagrangian, later spontaneously broken, but some fields are not invariant under $A_4$ transformations, much in the same way as the implementation of $SU(2)_L\times U(1)_Y$ in the SM, where left-handed and right-handed fermions are assigned to different representations of the gauge group.) Then we take the usual Higgs boson doublet $\Phi$ to be invariant under $A_4$, that is to be a flavor-singlet ${\bf 1}$ with no $T$-flavor. The other Higgs doublet $\eta$, the Higgs singlet $\chi$, and the singlet neutrinos $N_R$ are assumed to be triplets under $A_4$, and can so be used to introduce lepton-flavor violation in an $A_4$ symmetric Lagrangian.

The field content of our model and the field assignments to $SU(2)_L\times U(1)_Y\times A_{4}$ representations are summarized in Table~\ref{reps}.
These representation assignments and the requirement that the Lagrangian be renormalizable and $A_4$-symmetry forbid the presence of tree-level leptonic flavor-changing charged currents.

\begin{table}[b]
\begin{widetext}
\begin{center}
\caption{\label{reps} Representations of the fields under $A_4$ and $SU(2)_L \times U(1)_Y$. }
\begin{ruledtabular}
\begin{tabular}{ccccccccccccc}
Field &$L_{e},L_{\mu},L_{\tau}$&$e_R,\mu_R,\tau_R$&$N_{R}$&$\chi$&$\Phi$&$\eta$\\
\hline
$A_4$&$\mathbf{1}$, $\mathbf{1^\prime}$, $\mathbf{1^{\prime\prime}}$&$\mathbf{1}$, $\mathbf{1^\prime}$, $\mathbf{1^{\prime\prime}}$&$\mathbf{3}$&$\mathbf{3}$&$\mathbf{1}$&$\mathbf{3}$\\
$SU(2)_L\times U(1)_Y$&$(2,-\frac{1}{2})$&$(1,-1)$&$(1,0)$&$(1,0)$&$(2,\frac{1}{2})$&$(2,\frac{1}{2})$\\
\end{tabular}
\end{ruledtabular}
\end{center}
\end{widetext}
\end{table}

The renormalizable Yukawa interactions in the neutrino and charged lepton sectors invariant under $SU(2)_L\times U(1)_Y\times A_{4}$ are (including a Majorana mass term for the right-handed neutrinos)
 \begin{eqnarray}
 -{\cal L}_{\rm Yuk} &=& y^{\nu}_{1}\bar{L}_{e}(\tilde{\eta}N_{R})_{{\bf 1}}+y^{\nu}_{2}\bar{L}_{\mu}(\tilde{\eta}N_{R})_{{\bf 1}'}+y^{\nu}_{3}\bar{L}_{\tau}(\tilde{\eta}N_{R})_{{\bf 1}''}\nonumber\\
 &+&\frac{1}{2}M(\overline{N^{c}_{R}}N_{R})_{{\bf 1}}+\frac{1}{2}y_R^\nu(\overline{N^{c}_{R}}N_{R})_{{\bf 3}_{s}} \chi\nonumber\\
 &+& y_{e}\bar{L}_{e}\Phi~e_{R}+y_{\mu}\bar{L}_{\mu}\Phi~\mu_{R}+y_{\tau}\bar{L}_{\tau}\Phi~ \tau_{R}+\text{h.c.},
 \label{lagrangian}
 \end{eqnarray}
where $\tilde{\eta}\equiv i\tau_{2}\eta^{\ast}$ and $\tau_{2}$ is a Pauli matrix. In this Lagrangian, each flavor of neutrinos and each flavor of charged leptons has its own independent Yukawa term, since they belong to different singlet representations ${\bf 1}$, ${\bf 1}'$, and ${\bf 1}''$ of $A_4$: the neutrino Yukawa terms involve the $A_{4}$-triplets $\eta$ and $N_R$, which combine into the appropriate singlet representation; the charged-lepton Yukawa terms involve the $A_{4}$-singlet $\Phi$ and the $A_4$-singlet right-handed charged-leptons $e_R$, $\mu_R$, and $\tau_R$. The right-handed neutrinos have an additional Yukawa term that involves the $A_4$-triplet SM-singlet Higgs $\chi$.
The mass term $\frac{1}{2}M(\overline{N^{c}_{R}}N_{R})_{{\bf 1}}$ for the right-handed neutrinos is necessary to implement the seesaw mechanism by making the right-handed neutrino mass parameter $M$ large.

The Higgs potential of our model contains many terms and is listed in Appendix A, Eqs.~(\ref{potential1})--(\ref{potential6}).\footnote{We note that at TeV-scale the higher dimensional operators $(d\geq5)$ driven by $\chi$ and $\eta$ fields are suppressed by a cutoff scale $\Lambda$ which we assume is a very high energy scale, i.e.\ GUT or Planck scale. And in this paper we neglect the effects of higher dimensional operators.}
We spontaneously break the $A_4$ flavor symmetry by giving non-zero vacuum expectation values to some components of the $A_4$-triplets $\chi$ and $\eta$.
As seen in Appendix A, the minimization of our scalar potential gives the following vacuum expectation values (VEVs), all real:
 \begin{eqnarray}
\langle\varphi^{0}\rangle= \frac{v_{\Phi}}{\sqrt{2}}\neq0,
\quad
\langle\eta^{0}_{1}\rangle=\langle\eta^{0}_{2}\rangle=\langle\eta^{0}_{3}\rangle\equiv \frac{v_{\eta}}{\sqrt{2}}\neq0,
\quad
\langle\chi_{1}\rangle\equiv v_{\chi}\neq0~,\quad\langle\chi_{2}\rangle=\langle\chi_{3}\rangle=0.
 \label{subgroup}
 \end{eqnarray}
 The SM VEV $v=(\sqrt{2}G_{F})^{-1/2}=246$ GeV results from the combination $v=\sqrt{v^{2}_{\Phi}+3v^{2}_{\eta}}$.
The non-zero expectation value $\langle \varphi^0 \rangle = v_\Phi/\sqrt{2}$ does not break the $A_4$ symmetry, because the standard model Higgs is $A_4$-flavorless.
The non-zero expectation value $\langle \eta \rangle = ( v_\eta, v_\eta, v_\eta)/\sqrt{2}$ breaks the $S$-flavor-parity $(\eta_1,\eta_2,\eta_3) \to (\eta_1,-\eta_2,-\eta_3)$ but leaves the vacuum $T$-flavor  $T_f=0$. In other words, after $\eta$ acquires a non-zero VEV, the $T$-flavor is still conserved but the $S$-flavor-parity is not. Since $\eta$ appears only in the Higgs sector and in interactions with the light leptons, we say that the light neutrino sector has a residual $Z_3$ symmetry expressed by the subgroup $\{ 1, T, T^2 \}$ that leads to the conservation of $T$-flavor in terms involving mixing with the light neutrinos or interactions with the charged leptons.
The non-zero expectation value $\langle \chi \rangle = (v_\chi,0,0)$ maintains the $S$-flavor-parity of the vacuum (it is $S$-flavor-even) but gives the vacuum the symmetric combination of $T$-flavors $(a_0+a_{+1}+a_{-1})/\sqrt{3}$. That is, after $\chi$ acquires a non-zero VEV, the $S$-flavor-parity is  conserved but the $T$-flavor is not.  Since $\chi$ appears only in the Higgs sector and in interactions with the heavy Majorana neutrinos, we say that the heavy neutrino sector has a residual $Z_2$ symmetry expressed by the subgroup $\{1,S\}$ leading to the conservation of  $S$-flavor-parity in terms involving mixing or interactions with the heavy Majorana neutrinos.

When a non-Abelian discrete symmetry like our $A_4$ is considered, it is crucial to check the stability of the vacuum. In the presence of two $A_{4}$-triplet Higgs scalars $\chi$ and $\eta$, Higgs potential terms involving both $\chi$ and $\eta$, which would be written as $V(\chi\eta)$ in Eqs.~(\ref{potential1})--(\ref{potential6}), would be problematic for vacuum stability. Such stability problems can be naturally solved, for instance, in the presence of extra dimensions or in supersymmetric dynamical completions~\cite{A4, vacuum}. In these cases, $V(\chi\eta)$ is not allowed or highly suppressed.

The physical Higgs fields are obtained in the usual way. In the Higgs sector we have four Higgs doublets $\Phi$, $\eta_1$, $\eta_2$ and $\eta_3$, and three Higgs singlets $\chi_1$, $\chi_2$, and $\chi_3$. They contain in total 16 degrees of freedom: six charged Higgs fields $\eta^{\pm}_{1,2,3}$, with $\eta^{+}_{j}\equiv(\eta^{-}_{j})^{\ast}$, seven neutral Higgs scalars $h$, $h_{1,2,3}$, $\chi^{0}_{1,2,3}$, and three Higgs pseudoscalars $A_{1,2,3}$. We can write, after electroweak- and $A_4$-symmetry breaking and minimization of the potential,
 \begin{eqnarray}
  \Phi &=&
  {\left(\begin{array}{c}
  \varphi^{+} \\
  \frac{1}{\sqrt{2}}\left(v_{\Phi}+h+iA_{0}\right)
 \end{array}\right)}~,\quad\chi_{1}=v_{\chi}+\chi^{0}_{1}~,\quad \chi_{2}=\chi^{0}_{2}~,\quad \chi_{3}=\chi^{0}_{3}~,\nonumber\\
 \eta_{j} &=&
  {\left(\begin{array}{c}
  \eta^{+}_{j} \\
  \frac{1}{\sqrt{2}}\left(v_{\eta}+h_{j}+iA_{j}\right)
 \end{array}\right)}~,\quad j=1,2,3.
  \label{Higgs_vev}
 \end{eqnarray}
The action of the residual $Z_2$ generator $S$ on the physical fields is
\begin{align}
\label{eq:S1}
& (N_{R1}, N_{R2}, N_{R3} ) \to (N_{R1}, -N_{R2}, -N_{R3}) , \\
& (\chi^0_1, \chi^0_2, \chi^0_3) \to (\chi^0_1, -\chi^0_2, -\chi^0_3 ) , \\
& (h_{1}, h_{2}, h_{3} ) \to (h_{1}, -h_{2}, -h_{3}) , \\
& (A_{1}, A_{2}, A_{3} ) \to (A_{1}, -A_{2}, -A_{3}) , \\
& (\eta^{+}_{1}, \eta^{+}_{2}, \eta^{+}_{3} ) \to (\eta^{+}_{1}, -\eta^{+}_{2}, -\eta^{+}_{3}) ,
\label{eq:S2}
\end{align}
all other fields are invariant.
The action of the residual $Z_3$ generator $T$ on the physical fields is (the triplet fields $a_1$, $a_2$, and $a_3$ and the triplet fields $a_e$, $a_\mu$, and $a_\tau$ are linear combinations of each other, see Eqs.~(\ref{eq:Ua1})--(\ref{eq:Ua2}))
\begin{eqnarray}
\label{eq:T1}
%& (N_{R1}, N_{R2}, N_{R3} ) \to (N_{R2}, N_{R3}, N_{R1}) , \\
%& (h_{1}, h_{2}, h_{3} ) \to (h_{2}, h_{3}, h_{1}) , \\
%& (A_{1}, A_{2}, A_{3} ) \to (A_{2}, A_{3}, A_{1}) , \\
& (e,\mu,\tau) \to (e,\omega \mu, \omega^2 \tau) , \\
& (\nu_e,\nu_\mu,\nu_\tau) \to (\nu_e,\omega \nu_\mu, \omega^2 \nu_\tau) , \\
& (N_{Re},N_{R\mu},N_{R\tau}) \to (N_{Re},\omega N_{R\mu}, \omega^2 N_{R\tau}) , \\
& (\chi^0_e, \chi^0_\mu, \chi^0_\tau) \to (\chi^0_e, \omega \chi^0_\mu, \omega^2 \chi^0_\tau ) , \\
& (h_e,h_\mu,h_\tau) \to (h_e,\omega h_\mu, \omega^2 h_\tau) , \\
& (A_e,A_\mu,A_\tau) \to (A_e,\omega A_\mu, \omega^2 A_\tau) , \\
& (\eta^{+}_e,\eta^{+}_\mu,\eta^{+}_\tau) \to (\eta^{+}_e,\omega \eta^{+}_\mu, \omega^2 \eta^{+}_\tau) ,
\label{eq:T2}
\label{int}
\end{eqnarray}
all other fields are invariant.

After electroweak and $A_4$ symmetry breaking, the neutral Higgs fields acquire vacuum expectation values and give masses to the charged-leptons and neutrinos: the Higgs doublet gives Dirac masses to the charge leptons, the Higgs doublet $\eta$ gives Dirac masses to the three SM neutrinos, and the Higgs singlet $\chi$ gives a Majorana mass to the right-handed neutrino $N_R$.

The charged lepton mass matrix is automatically diagonal due to the $A_{4}$-singlet nature of the charged lepton and SM-Higgs fields. The right-handed neutrino mass has the (large) Majorana mass contribution  $M$  and a contribution  induced by the electroweak-singlet $A_4$-triplet Higgs boson $\chi$ when the $A_4$-symmetry is spontaneously broken.

After the breaking of the flavor and electroweak symmetries, with the VEV alignments as in Eq.~(\ref{subgroup}), the charged lepton, Dirac neutrino and right-handed neutrino mass terms from the Lagrangian~(\ref{lagrangian}) result in
 \begin{eqnarray}
 -{\cal L}_{m} &=& \frac{v_{\Phi}}{\sqrt{2}}\left(y_{e}\bar{e}_{L}e_{R}+y_{\mu}\bar{\mu}_{L}\mu_{R}+y_{\tau}\bar{\tau}_{L}\tau_{R}\right)+\frac{v_{\eta}}{\sqrt{2}}\Big\{\left(y^{\nu}_{1}\bar{\nu}_{e}+y^{\nu}_{2}\bar{\nu}_{\mu}+y^{\nu}_{3}\bar{\nu}_{\tau}\right)N_{R1}\nonumber\\
 &+&\left(y^{\nu}_{1}\bar{\nu}_{e}+y^{\nu}_{2}\omega\bar{\nu}_{\mu}+y^{\nu}_{3}\omega^{2}\bar{\nu}_{\tau}\right)N_{R2}+\left(y^{\nu}_{1}\bar{\nu}_{e}+y^{\nu}_{2}\omega^{2}\bar{\nu}_{\mu}+y^{\nu}_{3}\omega\bar{\nu}_{\tau}\right)N_{R3}\Big\}\nonumber\\
 &+&\frac{M}{2}(\overline{N^{c}_{R1}}N_{R1}+\overline{N^{c}_{R2}}N_{R2}+\overline{N^{c}_{R3}}N_{R3})
 +\frac{y_R^\nu v_{\chi}}{2}(\overline{N^{c}_{R2}}N_{R3}+\overline{N^{c}_{R3}}N_{R2})+\text{h.c.}~.
 \label{lagrangian1}
 \end{eqnarray}
This form shows clearly that the terms in $v_\eta$ break the $S$-flavor-parity symmetry (\ref{eq:S1})--(\ref{eq:S2}), while the other mass terms preserve it. Passing to the $T$-flavor eigenfields
 \begin{align}
 N_{Re} & = \frac{N_{R1}+N_{R2}+N_{R3}}{\sqrt{3}} , \\
 N_{R\mu} & = \frac{N_{R1}+\omega N_{R2}+\omega^2 N_{R3}}{\sqrt{3}} , \\
 N_{R\tau} & = \frac{N_{R1}+\omega^2 N_{R2}+\omega N_{R3}}{\sqrt{3}} ,
 \end{align}
 with respective $T$-flavor $T_f=0,+1,-1$,
the lepton mass Lagrangian reads
 \begin{eqnarray}
 -{\cal L}_{m} &=& \frac{v_{\Phi}}{\sqrt{2}}\left(y_{e}\bar{e}_{L}e_{R}+y_{\mu}\bar{\mu}_{L}\mu_{R}+y_{\tau}\bar{\tau}_{L}\tau_{R}\right)\nonumber \\ &+&v_{\eta} \sqrt{\frac{3}{2}} \left(y^{\nu}_{1}\bar{\nu}_{e} N_{Re}+y^{\nu}_{2}\bar{\nu}_{\mu} N_{R\mu}+y^{\nu}_{3}\bar{\nu}_{\tau} N_{R\tau} \right)\nonumber\\[1ex]
 &+&\frac{M}{2}(\overline{N^{c}_{Re}}N_{Re}+\overline{N^{c}_{R\mu}}N_{R\tau}+\overline{N^{c}_{R\tau}}N_{R\mu})\nonumber\\[1ex]
 &+&\frac{y_R^\nu v_{\chi}}{2} \Big[\overline{N^{c}_{Re}}N_{Re}+\overline{N^{c}_{R\mu}}N_{R\mu}+\overline{N^{c}_{R\tau}}N_{R\tau}\nonumber\\&&\qquad\;-\frac{1}{3} \left( \overline{N^{c}_{Re}} + \overline{N^{c}_{R\mu}} + \overline{N^{c}_{R\tau}} \right) \left( N_{Re} + N_{R\mu} + N_{R\tau}\right)\Big]+\text{h.c.}~.
 \label{lagrangian1p}
 \end{eqnarray}
This form shows clearly that the terms in $v_\chi$ break the $T$-flavor symmetry (\ref{eq:T1})--(\ref{eq:T2}), while the other mass terms preserve it.

Inspection of the mass terms in Eq.~(\ref{lagrangian1p}) indicates that, with the VEV alignments in Eq.~(\ref{subgroup}), the $A_{4}$ symmetry is spontaneously broken to a residual $Z_{2}$ symmetry in the heavy Majorana neutrino sector (conservation of $S$-flavor-parity in terms not involving $v_\eta$ or $h_{1,2,3}$) and a residual $Z_{3}$ symmetry in the Dirac neutrino sector (conservation of $T$-flavor in terms not involving $v_\chi$ or $\chi^0_1$).

%It is convenient to classify all the particles by $Z_{3}$ charges through performing the unitary transformation $T\rightarrow T'=U_{\omega}TU^{\dag}_{\omega}$ where $U_{\omega}$ is defined in Eq.~(\ref{YnuT}). The transformations of fields under $Z_3$ symmetry are given by
%\begin{align}
%& N'_{R1} \to N'_{R1} , && N'_{R2} \to \omega N'_{R2}, && N'_{R3} \to \omega^2 N'_{R3} \\
%& h'_1 \to h'_1 , && h'_2 \to \omega h'_2 , && h'_3 \to \omega^2 h'_3 , \\
%& A'_1 \to h'_1 , && A'_2 \to \omega A'_2 , && A'_3 \to \omega^2 A'_3 , \\
%& \nu_e \to \nu_e, && \nu_\mu \to \omega \nu_\mu, && \nu_\tau \to \omega^2 \nu_\tau , \\
%& e \to e, && \mu \to \omega \mu, && \tau \to \omega^2 \tau .
%\label{int2}
%\end{align}
%The residual $Z_{3}$ $T$-flavor symmetry has not yet been observed in Nature.
%It may give rise to a stable dark matter candidate, as the lightest combination of
%$h_{\mu}$, $h_{\tau}$, $A_{\mu}$ and $A_{\tau}$, which couples only to the heavy
%right-handed neutrinos and not to the SM charged fermions~\cite{A4DM}.

The mass terms in Eq.~(\ref{lagrangian1}) and the charged gauge interactions in the weak eigenstate basis can be written in (block) matrix form as, using $\overline{N^{c}_R} m_D \nu^{c}_L = \overline{\nu_L} m_D^T N_R $,
 \begin{eqnarray}
 -{\cal L}_{mW} &=& \frac{1}{2}\overline{N^{c}_{R}}M_{R}N_{R}
 +\overline{\nu_{L}}m_{D}N_{R}+\overline{\ell_{L}}m_{\ell}\ell_{R}+\frac{g}{\sqrt{2}}W^{-}_{\mu}\overline{\ell_{L}}\gamma^{\mu}\nu_{L}+\text{h.c.}~ \\
 &=& \frac{1}{2} \begin{pmatrix} \overline{\nu_L} & \overline{N^{c}_R} \end{pmatrix} \begin{pmatrix} 0 & m_D \\ m_D^T & M_R \end{pmatrix} \begin{pmatrix} \nu^{c}_L \\ N_R \end{pmatrix} + \overline{\ell_{L}}m_{\ell}\ell_{R}+\frac{g}{\sqrt{2}}W^{-}_{\mu}\overline{\ell_{L}}\gamma^{\mu}\nu_{L}+\text{h.c.}
 \label{lagrangianA}
 \end{eqnarray}
Here $\ell=(e,\mu,\tau)$, $\nu=(\nu_e,\nu_\mu,\nu_\tau)$, $N_R=(N_{R1},N_{R2},N_{R3})$, and
\begin{align}
m_{\ell} & =\frac{v_{\Phi}}{\sqrt{2}} \begin{pmatrix} y_{e} & 0 & 0 \\ 0 & y_{\mu} & 0 \\ 0 & 0 & y_{\tau}\end{pmatrix} , \\
m_{D}& =\frac{v_{\eta}}{\sqrt{2}} \, Y_{\nu} = \frac{v_\eta}{\sqrt{2}} \begin{pmatrix}
y^{\nu}_1 & y^{\nu}_1 & y^{\nu}_1 \\
y^{\nu}_2 & \omega y^{\nu}_2 & \omega^2 y^{\nu}_2 \\
y^{\nu}_3 & \omega^2 y^{\nu}_3 & \omega y^{\nu}_3
\end{pmatrix} ,
\label{eq:Ynu}
\\
M_R & =  \begin{pmatrix} M & 0 & 0 \\ 0 & M & y^{\nu}_R v_\chi \\ 0 & y^{\nu}_R v_\chi & M \end{pmatrix} .
\end{align}

To find the neutrino masses and mixing matrix we are to diagonalize the $6\times6$ matrix
\begin{align}
 \begin{pmatrix} 0 & m_D \\ m^{T}_D & M_R \end{pmatrix} .
\end{align}

We start by diagonalizing $M_R$. For this purpose, we perform a basis rotation $\widehat{N}_{R} = U^{\dag}_{R}N_{R}$,
so that the right-handed Majorana mass matrix $M_{R}$ becomes a diagonal matrix $\widehat{M}_R$ with real and positive mass eigenvalues $M_{1}=a M$, $M_{2}=M$ and $M_{3}=b M$,
 \begin{eqnarray}
  \widehat{M}_{R}&=& U^{T}_{R}M_{R}U_{R}=MU^{T}_{R}{\left(\begin{array}{ccc}
 1 &  0 &  0 \\
 0 &  1 &  \kappa e^{i\xi} \\
 0 &  \kappa e^{i\xi} &  1
 \end{array}\right)}U_{R}
= \begin{pmatrix} aM & 0 & 0 \\ 0 & M & 0 \\ 0 & 0 & bM \end{pmatrix},
  \label{heavy}
 \end{eqnarray}
where $\kappa=|y_R^\nu v_{\chi}/M|$ and $\xi=\arg(y_R^\nu v_{\chi}/M)$. We find $a=\sqrt{1+\kappa^{2}+2\kappa\cos\xi}$, $b=\sqrt{1+\kappa^{2}-2\kappa\cos\xi}$, and a diagonalizing matrix
\begin{eqnarray}
  U_{R} = \frac{1}{\sqrt{2}}{\left(\begin{array}{ccc}
  0  &  \sqrt{2}  &  0 \\
  1 &  0  &  -1 \\
  1 &  0  &  1
  \end{array}\right)}{\left(\begin{array}{ccc}
  e^{i\frac{\psi_1}{2}}  &  0  &  0 \\
  0  &  1  &  0 \\
  0  &  0  &  e^{i\frac{\psi_2}{2}}
  \end{array}\right)}~,
  \label{URN}
\end{eqnarray}
with phases
\begin{eqnarray}
 \psi_1 = \tan^{-1} \Big( \frac{-\kappa\sin\xi}{1+\kappa\cos\xi} \Big)
 ~~~{\rm and}~~~ \psi_2 = \tan^{-1} \Big( \frac{\kappa\sin\xi}{1-\kappa\cos\xi} \Big)~.
\label{alphs_beta}
\end{eqnarray}
As the magnitude of $\kappa$ defined in Eq.~(\ref{heavy}) decreases, the phases $\psi_{1,2}$ go to $0$ or $\pi$. At this point,
\begin{align}
-{\cal L}_{mW} &= \frac{1}{2} \begin{pmatrix} \overline{\nu_L} & \overline{\widehat{N}^{c}_R} \end{pmatrix} \begin{pmatrix} 0 & \widetilde{m}_D \\ \widetilde{m}_D^T & \widehat{M}_R \end{pmatrix} \begin{pmatrix} \nu^{c}_L \\ \widehat{N}_R \end{pmatrix} + \overline{\ell_{L}}m_{\ell}\ell_{R}+\frac{g}{\sqrt{2}}W^{-}_{\mu}\overline{\ell_{L}}\gamma^{\mu}\nu_{L}+\text{h.c.}
 \label{lagrangianB}
\end{align}
with $\widetilde{m}_D = m_D U_R$.

Now we take the limit of large $M$ (seesaw mechanism)  and focus on the mass matrix of the light neutrinos $M_{\nu}$,
\begin{align}
-{\cal L}_{mW} &= \frac{1}{2} \overline{\nu_L} M_{\nu} \nu^{c}_L  + \overline{\ell_{L}}m_{\ell}\ell_{R}+\frac{g}{\sqrt{2}}W^{-}_{\mu}\overline{\ell_{L}}\gamma^{\mu}\nu_{L}+\text{h.c.}+\text{terms in $N_R$}
\end{align}
with
 \begin{align}
 M_{\nu} = - \widetilde{m}_D \, \widehat{M}_R^{-1} \, \widetilde{m}^T_D.
 \end{align}
 We perform basis rotations from weak  to mass eigenstates in the leptonic sector,
\begin{eqnarray}
 \widehat{\ell}_{L} = P^{\ast}_{\ell}\ell_{L}~,\quad \widehat{\ell}_{R}= P^{\ast}_{\ell}\ell_{R}~,\quad \widehat{\nu}_{L} = U^{\dag}_{\nu}P^{\ast}_{\nu}\nu_{L}~,
 \label{rebasing}
\end{eqnarray}
where $P_{\ell}$ and $P_{\nu}$ are phase matrices and $U_\nu$ is a unitary matrix chosen so as the matrix
\begin{align}
\widehat{m}_{\nu} = U^{\dag}_{\nu} P_\nu^* M_\nu P_\nu^* U^*_{\nu} = - U_{\nu}^{\dag}P_{\nu}^{\ast} m_D U_R   \widehat{M}_R^{-1} (U_{\nu}^{\dag}P_{\nu}^{\ast} m_D U_R)^T
\end{align}
is diagonal.
Then from the charged current term in Eq.~(\ref{lagrangianB}) we obtain the lepton mixing matrix $U_{\rm PMNS}$ as
\begin{align}
U_{\rm PMNS}=P^{\ast}_{\ell}P_{\nu}U_{\nu}.
\end{align}
The matrix $U_{\rm PMNS}$ can be written in terms of three mixing angles and three $CP$-odd phases (one for the Dirac neutrinos and two for the Majorana neutrinos) as \cite{PDG}
\begin{eqnarray}
  U_{\rm PMNS}
  &=&{\left(\begin{array}{ccc}
   c_{13}c_{12} & c_{13}s_{12} & s_{13}e^{-i\delta_{CP}} \\
   -c_{23}s_{12}-s_{23}c_{12}s_{13}e^{i\delta_{CP}} & c_{23}c_{12}-s_{23}s_{12}s_{13}e^{i\delta_{CP}} & s_{23}c_{13}  \\
   s_{23}s_{12}-c_{23}c_{12}s_{13}e^{i\delta_{CP}} & -s_{23}c_{12}-c_{23}s_{12}s_{13}e^{i\delta_{CP}} & c_{23}c_{13}
   \end{array}\right)}Q_{\nu}~,
 \label{rebasing1}
\end{eqnarray}
where $Q_{\nu}={\rm Diag}(e^{-i\varphi_{1}/2},e^{-i\varphi_{2}/2},1)$, and $s_{ij}\equiv \sin\theta_{ij}$ and $c_{ij}\equiv \cos\theta_{ij}$.

It is important to notice that the phase matrix $P_{\nu}$ can be rotated away by choosing the matrix $P_{\ell}=P_\nu$, i.e.\ by an appropriate redefinition of the left-handed charged lepton fields, which is always possible. This is an important point because the phase matrix $P_{\nu}$ accompanies the Dirac-neutrino mass matrix $\tilde{m}_D$ and ultimately the neutrino Yukawa matrix $Y_{\nu}$ in Eq.~(\ref{eq:Ynu}). This means that complex phases in $Y_{\nu}$ can always be rotated away by appropriately choosing the phases of left-handed charged lepton fields. Hence without loss of generality the eigenvalues $y^{\nu}_1$, $y^{\nu}_2$, and $y^{\nu}_3$ of $Y_{\nu}$ can be real and positive.
The Yukawa matrix $Y_{\nu}$ can then be written as
 \begin{eqnarray}
 Y_{\nu}=y^{\nu}_{3}\sqrt{3}{\left(\begin{array}{ccc}
 y_{1} &  0 &  0 \\
 0 & y_{2} & 0 \\
 0 & 0 & 1
 \end{array}\right)}U^{\dag}_{\omega},
  \label{YnuT}
 \end{eqnarray}
where $y_{1}=|y^{\nu}_{1}/y^{\nu}_{3}|, y_{2}=|y^{\nu}_{2}/y^{\nu}_{3}|$, and $U_\omega$ is given in Eq.~(\ref{eq:Uomega}).

Concerning CP violation, we notice that the CP phases $\psi_1,\psi_2$ coming from $M_{R}$ only take part in low-energy CP violation, as can be seen in Eqs.~(\ref{heavy}-\ref{YnuT}). Any CP-violation relevant for leptogenesis is associated with the neutrino Yukawa matrix $\widetilde{Y}_{\nu}=Y_{\nu}U_{R}$ and the combination of Dirac neutrino Yukawa matrices, $H\equiv\widetilde{Y}^{\dag}_{\nu}\widetilde{Y}_{\nu}= U^{\dag}_{R}Y^{\dag}_{\nu}Y_{\nu}U_{R}$, which is
 \begin{eqnarray}
   H=3|y^{\nu}_{3}|^{2}\left(\begin{array}{ccc}
  \frac{1+4y^{2}_{1}+y^{2}_{2}}{2} & \frac{e^{-i\frac{\psi_{1}}{2}}}{\sqrt{2}}(2y^{2}_{1}-y^{2}_{2}-1) & \frac{i\sqrt{3}e^{i\frac{\psi_{21}}{2}}}{2}(y^{2}_{2}-1) \\
  \frac{e^{i\frac{\psi_{1}}{2}}}{\sqrt{2}}(2y^{2}_{1}-y^{2}_{2}-1) & 1+y^{2}_{1}+y^{2}_{2} & -i\sqrt{\frac{3}{2}}e^{i\frac{\psi_{2}}{2}}(y^{2}_{2}-1) \\
  -\frac{i\sqrt{3}e^{-i\frac{\psi_{21}}{2}}}{2}(y^{2}_{2}-1) & i\sqrt{\frac{3}{2}}e^{-i\frac{\psi_{2}}{2}}(y^{2}_{2}-1) & \frac{3}{2}(1+y^{2}_{2})
  \end{array}\right) ,
 \label{YnuYnu}
 \end{eqnarray}
where $\psi_{ij}\equiv\psi_{i}-\psi_{j}$. As expected, in the limit $|y^{\nu}_1|=|y^{\nu}_2|=|y^{\nu}_3|$ , i.e.\ $y_{1,2}\rightarrow1$, the off-diagonal entries of $H$  vanish, and there is no CP violation useful for leptogenesis. If the Dirac neutrino Yukawa couplings $y^{\nu}_1$, $y^{\nu}_2$, and $y^{\nu}_3$ differ in magnitude, they can play a role in baryogenesis via leptogenesis and non-zero $\theta_{13}\simeq9^{\circ}$ with TBM ($\theta_{23}\simeq45^{\circ},\theta_{12}\simeq35^{\circ}$). Therefore,
a low energy CP violation in neutrino oscillation and/or a high energy CP
violation in leptogenesis can be generated by the non-degeneracy of the Dirac neutrino Yukawa couplings and a non-zero phase $\xi$ coming from $M_{R}$.

In the following section we investigate the low energy phenomenology, namely the possible values of the light neutrino mixing angles, how the low energy CP violation could be generated in both normal and inverted mass hierarchies, and neutrinoless double beta decay, which is a probe of lepton number violation at low energy.

%%%%%%%%%%%%%%%%%%%%%%%%%%%%%%%%%%%%%%%%%%%%%%%%%%%%%%%%%%%%%%%%%%%%%%%%%%%%%%%%%%%%
\section{Phenomenology of light neutrinos}
After seesawing, in a basis where charged lepton and heavy neutrino masses are real and diagonal, the light neutrino mass matrix is given by
 \begin{eqnarray}
  m_{\nu} &=& -\widetilde{m}_{D}\widehat{M}^{-1}_{R}\widetilde{m}^{T}_{D} =-\frac{v^{2}_{\eta}}{2}Y_{\nu}U_{R}\widehat{M}^{-1}_{R}U^{T}_{R}Y^{T}_{\nu}\nonumber\\
   &=& m_{0}
   {\left(\begin{array}{ccc}
   (1+\frac{2e^{i\psi_{1}}}{a})y^{2}_{1} & (1-\frac{e^{i\psi_{1}}}{a})y_{1}y_{2} & (1-\frac{e^{i\psi_{1}}}{a})y_{1} \\
   (1-\frac{e^{i\psi_{1}}}{a})y_{1}y_{2} & (1+\frac{e^{i\psi_{1}}}{2a}-\frac{3e^{i\psi_{2}}}{2b})y^{2}_{2} & (1+\frac{e^{i\psi_{1}}}{2a}+\frac{3e^{i\psi_{2}}}{2b})y_{2}  \\
   (1-\frac{e^{i\psi_{1}}}{a})y_{1} & (1+\frac{e^{i\psi_{1}}}{2a}+\frac{3e^{i\psi_{2}}}{2b})y_{2} & (1+\frac{e^{i\psi_{1}}}{2a}-\frac{3e^{i\psi_{2}}}{2b})
   \end{array}\right)}~,
  \label{meff}
 \end{eqnarray}
where $\tilde{m}_{D}=v_{\eta}\widetilde{Y}_{\nu}/\sqrt{2}$ and we have defined an overall scale $m_{0}=v^{2}_{\eta}|y^{\nu2}_{3}|/(6M)$ for the light neutrino masses.
The mass matrix $m_\nu$ is diagonalized by the PMNS mixing matrix $U_{\rm PMNS}$ as described above,
 \begin{eqnarray}
  m_{\nu} &=& U_{\rm PMNS} ~{\rm Diag}(m_{1},m_{2},m_{3})~ U^{T}_{\rm PMNS} .
 \end{eqnarray}
Here $m_i$ $(i = 1,2,3)$ are the light neutrino masses. As is well-known, because of the observed hierarchy $|\Delta m^{2}_{\rm Atm}|\equiv |m^{2}_{3}-m^{2}_{1}|\gg\Delta m^{2}_{\rm Sol}\equiv m^{2}_{2}-m^{2}_{1}>0$, and the requirement of a Mikheyev-Smirnov-Wolfenstein resonance for solar neutrinos, there are two possible neutrino mass spectra: (i) the normal mass hierarchy (NMH) $m_{1}<m_{2}<m_{3}$, and (ii) the inverted mass hierarchy (IMH) $m_{3}<m_{1}<m_{2}$.

Interestingly, the combination $U^{\dag}_{\omega}U_{R}$ in Eq.~(\ref{meff}) reflects an exact TBM:
 \begin{eqnarray}
 U^{\dag}_{\omega}U_{R}={\left(\begin{array}{ccc}
 \sqrt{\frac{2}{3}} &  \frac{1}{\sqrt{3}}&  0 \\
 -\frac{1}{\sqrt{6}}& \frac{1}{\sqrt{3}} & -\frac{1}{\sqrt{2}} \\
 -\frac{1}{\sqrt{6}}& \frac{1}{\sqrt{3}} &  \frac{1}{\sqrt{2}}
 \end{array}\right)}{\left(\begin{array}{ccc}
  e^{i\frac{\psi_1}{2}}  &  0  &  0 \\
  0  &  1  &  0 \\
  0  &  0  &  e^{i\frac{\psi_2+\pi}{2}}
  \end{array}\right)}~.
 \label{UwUr}
 \end{eqnarray}
Therefore Eq.~(\ref{meff}) directly indicates that there could be deviations from the exact TBM if the Dirac neutrino Yukawa couplings do not have the same magnitude.
In the limit $|y^\nu_2|=|y^\nu_3|$ ($y_{2}\rightarrow1$), the mass matrix in Eq.~(\ref{meff}) acquires a $\mu$--$\tau$ symmetry that leads to $\theta_{13}=0$ and $\theta_{23}=-\pi/4$. Moreover, in the limit  $|y^\nu_1|=|y^\nu_2|=|y^\nu_3|$ ($y_{1}, y_{2}\rightarrow1$), the  mass matrix~(\ref{meff}) gives the TBM angles in Eq.~(\ref{TBM}) and the corresponding mass eigenvalues
 \begin{eqnarray}
 m_{1}&=& \frac{3m_{0}}{a}~,\qquad m_{2}=3m_{0}~,\qquad m_{3}= \frac{3m_{0}}{b}~.
 \label{TBM1}
 \end{eqnarray}
 These mass eigenvalues are disconnected from the mixing angles. However, recent neutrino data, {\it i.e.} $\theta_{13}\neq0$, require deviations of $y_{1,2}$ from unity, leading to a possibility to search for $CP$ violation in neutrino oscillation experiments.
These deviations generate relations between mixing angles and mass eigenvalues.

To diagonalize the above mass matrix Eq.~(\ref{meff}), we consider the hermitian matrix $m_{\nu}m^{\dag}_{\nu}=U_{\rm PMNS}~{\rm Diag}(m^{2}_{1},m^{2}_{2},m^{2}_{3})~U^{\dag}_{\rm PMNS}$, from which we obtain the masses and mixing angles.
To see how the neutrino mass matrix given by Eq.(\ref{meff}) can lead to deviations of neutrino mixing angles from their TBM values,
 we first introduce three small quantities $\epsilon_{i},~(i=1,2,3)$, which are responsible for the deviations of the $\theta_{jk}$ from their TBM values:
 \begin{eqnarray}
  \theta_{23}=-\frac{\pi}{4}+\epsilon_{1}~, \qquad\theta_{13}=\epsilon_{2}~, \qquad\theta_{12}= \sin^{-1}\left(\frac{1}{\sqrt{3}}\right)+\epsilon_{3}~.
 \end{eqnarray}
Then the PMNS mixing matrix up to order $\epsilon_{i}$ can be written as
 \begin{eqnarray}
 U_{\rm PMNS}&=&{\left(\begin{array}{ccc}
 \frac{\sqrt{2}-\epsilon_{3}}{\sqrt{3}} &  \frac{1+\epsilon_{3}\sqrt{2}}{\sqrt{3}} &  \epsilon_{2}e^{-i\delta_{CP}} \\
 -\frac{1+\epsilon_{1}+\epsilon_{3}\sqrt{2}}{\sqrt{6}}+\frac{\epsilon_{2}e^{i\delta_{CP}}}{\sqrt{3}} &  \frac{\sqrt{2}+\epsilon_{1}\sqrt{2}-\epsilon_{3}}{\sqrt{6}}+\frac{\epsilon_{2}e^{i\delta_{CP}} }{\sqrt{6}} &  \frac{-1+\epsilon_{1}}{\sqrt{2}} \\
 \frac{-1+\epsilon_{1}+\epsilon_{3}\sqrt{2}}{\sqrt{6}}-\frac{\epsilon_{2}}{\sqrt{3}}e^{i\delta_{CP}} &  \frac{\sqrt{2}-\epsilon_{3}-\sqrt{2}\epsilon_{1}}{\sqrt{6}}-\frac{\epsilon_{2}}{\sqrt{6}}e^{i\delta_{CP}} &  \frac{1+\epsilon_{1}}{\sqrt{2}}
 \end{array}\right)}Q_{\nu}
 +{\cal O}(\epsilon^{2}_{i})~.
 \label{Unu}
 \end{eqnarray}

The small deviation $\epsilon_{1}$ from the maximality of the atmospheric mixing angle $\theta_{23}$ is expressed in terms of the parameters in Eq.~(\ref{MMele}) in Appendix B as
 \begin{eqnarray}
  \tan\epsilon_{1}=\frac{R(1+y_{2})-S(y_{2}-1)}{R(y_{2}-1)-S(1+y_{2})}~.
 \label{Atmdevi}
 \end{eqnarray}
In the limit of $y_{1},y_{2}\rightarrow1$, $\epsilon_1$ goes to zero (or equivalently $\theta_{23}\rightarrow-\pi/4$) due to $R,S\rightarrow0$.
The reactor angle $\theta_{13}$ and the Dirac-CP phase $\delta_{CP}$ are expressed as
 \begin{eqnarray}
  \tan2\theta_{13}&=&\frac{y_{1}|s_{23}(3Q-P)y_{2}-c_{23}(3Q+P)-3i\{s_{23}(R-S)y_{2}+c_{23}(R+S)\}|}{(F+G+\frac{9K}{4}+\frac{3D}{2})(c^{2}_{23}+y^{2}_{2}s^{2}_{23})+y_{2}(F+G-\frac{9K}{4})\sin2\theta_{23}-y^{2}_{1}\tilde{A}}~,\nonumber\\
  \tan\delta_{CP}&=&3\frac{(R-S)^{2}+y^{2}_{2}(R+S)^{2}}{(P+Q)(R-S)-y^{2}_{2}(P-Q)(R+S)}~,
 \label{DiracCP}
 \end{eqnarray}
where the parameters $P,Q,F,G,K,D$ and $\tilde{A}$ are given in Eq.~(\ref{MMele}) in Appendix B.
In the limit of  $y_{1},y_{2}\rightarrow1$, the parameters $Q,R,S$ go to zero, which in turn leads to  $\theta_{13}\rightarrow0$ and $\delta_{CP}\rightarrow0$ as expected.
Finally, the solar mixing angle is given by
 \begin{eqnarray}
  \tan2\theta_{12}=2y_{1}\frac{y_{2}c_{23}(3Q-P)+s_{23}(3Q+P)}{c_{13}(\Psi_{2}-\Psi_{1})}~.
 \label{sol1}
 \end{eqnarray}
Since  in the limit $y_{1},y_{2}\rightarrow1$ the parameters in Eq.~(\ref{sol1}) behave as $Q\rightarrow0,P\rightarrow6(\frac{1}{a^{2}}-1),\Psi_{1}\rightarrow3(1+\frac{2}{a^{2}})$ and $\Psi_{2}\rightarrow6(1+\frac{1}{2a^{2}})$, it is clear that the mixing angle $\tan2\theta_{12}$ goes to $2\sqrt{2}$, that is, $\theta_{12}\rightarrow\sin^{-1}(1/\sqrt{3})$.

The squared-mass eigenvalues of the three light neutrinos result in
 \begin{eqnarray}
    m^{2}_{1}&=&m^{2}_{0}\left\{s^{2}_{12}\Psi_{1}+c^{2}_{12}\Psi_{2}-y_{1}\frac{y_{2}c_{23}(3Q-P)+s_{23}(3Q+P)}{2c_{13}}\sin2\theta_{12}\right\}~,\nonumber\\
    m^{2}_{2}&=&m^{2}_{0}\left\{c^{2}_{12}\Psi_{1}+s^{2}_{12}\Psi_{2}+y_{1}\frac{y_{2}c_{23}(3Q-P)+s_{23}(3Q+P)}{2c_{13}}\sin2\theta_{12}\right\}~,\nonumber\\
    m^{2}_{3}&=&m^{2}_{0}\Big\{\Big[\left(F+G+\frac{9K}{4}+\frac{3D}{2}\right)(c^{2}_{23}+y^{2}_{2}s^{2}_{23})+y_{2}\left(F+G-\frac{9K}{4}\right)\sin2\theta_{23}\Big]c^{2}_{13}\nonumber\\
    &+&y^{2}_{1}\tilde{A}s^{2}_{13}-\frac{y_{1}\sin2\theta_{13}}{2}\Big[c_{23}\left((3Q+P)\cos\delta_{CP}-3(R+S)\sin\delta_{CP}\right)\nonumber\\
    &+&s_{23}y_{2}\left((3Q-P)\cos\delta_{CP}+3(R-S)\sin\delta_{CP}\right)\Big]\Big\}~.
 \label{eigenvalueGen}
 \end{eqnarray}
We see from Eqs.~(\ref{sol1}) and~(\ref{eigenvalueGen}) that the deviation $\epsilon_{3}$ from tri-maximality of solar mixing angle $\theta_{12}$ can be expressed as
 \begin{eqnarray}
  2\sqrt{2}\cos2\epsilon_{3}+\sin2\epsilon_{3}&=&\frac{3y_{1}m^{2}_{0}\{y_{2}c_{23}(3Q-P)+s_{23}(3Q+P)\}}{c_{13}\Delta m^{2}_{21}}~.
 \end{eqnarray}

Now we perform a numerical analysis using the linear algebra tools in Ref.~\cite{Antusch:2005gp}.
The Daya Bay and RENO experiments have
accomplished the measurement of three mixing angles
$\theta_{12}, \theta_{23}$, and $\theta_{13}$ from three kinds of neutrino oscillation experiments.
A combined analysis of the data from the T2K, MINOS, Double Chooz, Daya Bay and RENO experiments shows~\cite{Tortola:2012te} that, for the normal mass hierarchy (NMH) and inverted mass hierarchy (IMH), respectively,
\begin{eqnarray}
\sin^2\theta_{13}=0.026^{+0.003~(+0.010)}_{-0.004~(-0.011)}~{\rm NMH},\quad\left[0.027^{+0.003~(+0.010)}_{-0.004~(-0.011)}~{\rm IMH}\right]
\end{eqnarray}
or equivalently
\begin{eqnarray}
\theta_{13}=9.28^{\circ+0.53^{\circ}~(+1.66^{\circ})}_{~-0.75^{\circ}~(-2.24^{\circ})}~{\rm NMH},\quad\left[9.46^{+0.52^{\circ}~(+1.64^{\circ})}_{-0.73^{\circ}~(-2.19^{\circ})}~{\rm IMH}\right]
 \label{expdata}
\end{eqnarray}
at the $1\sigma~(3\sigma)$ level. The hypothesis $\theta_{13}=0$ is now rejected at the $8\sigma$ significance level.
In addition to the measurement of the mixing angle $\theta_{13}$, the global fit of the neutrino mixing angles and of the mass-squared differences at the $1\sigma$ $(3\sigma)$ level is
given by~\cite{Tortola:2012te}
 \begin{eqnarray}
  &&\theta_{12}=34.45^{\circ+0.92^{\circ}~(+3.02^{\circ})}_{~-1.05^{\circ}~(-3.14^{\circ})}~,\quad
  \theta_{23}=44.43^{\circ+4.60^{\circ}~(+8.70^{\circ})}_{~-2.87^{\circ}~(-5.78^{\circ})}~{\rm NMH},~\left[46.72^{\circ+2.89^{\circ}~(+6.41^{\circ})}_{~-4.01^{\circ}~(-8.07^{\circ})}~{\rm IMH}\right]
  \nonumber\\
  &&\Delta m^{2}_{\rm Sol}[10^{-5}{\rm eV}^{2}]=7.62^{+0.19~(+0.58)}_{-0.19~(-0.50)}~,\quad\Delta m^{2}_{\rm Atm}[10^{-3}{\rm eV}^{2}]=\left\{\begin{array}{ll}
                2.53^{+0.08~(+0.24)}_{-0.10~(-0.27)}~, & \hbox{NMH} \\
                2.40^{+0.10~(+0.28)}_{-0.07~(-0.25)}~, & \hbox{IMH}
                                  \end{array}
                                \right.
 \label{expnu}
 \end{eqnarray}

The matrices $m_{D}$ and $\hat{M}_{R}$ in Eq.~(\ref{meff}) contain seven parameters : $y^{\nu}_{3},M,v_{\eta},y_{1},y_{2},\kappa,\xi$. The first three ($y^{\nu}_{3}$, $M,$ and $v_{\eta}$) lead to the overall neutrino scale parameter $m_{0}$. The next four ($y_1,y_2,\kappa,\xi$) give rise to the deviations from TBM as well as the CP phases and corrections to the mass eigenvalues (see Eq.~(\ref{TBM1})).

\begin{figure}[t]
%\vspace*{-5.0cm}
%\hspace*{-1cm}
\begin{minipage}[t]{6.0cm}
\epsfig{figure=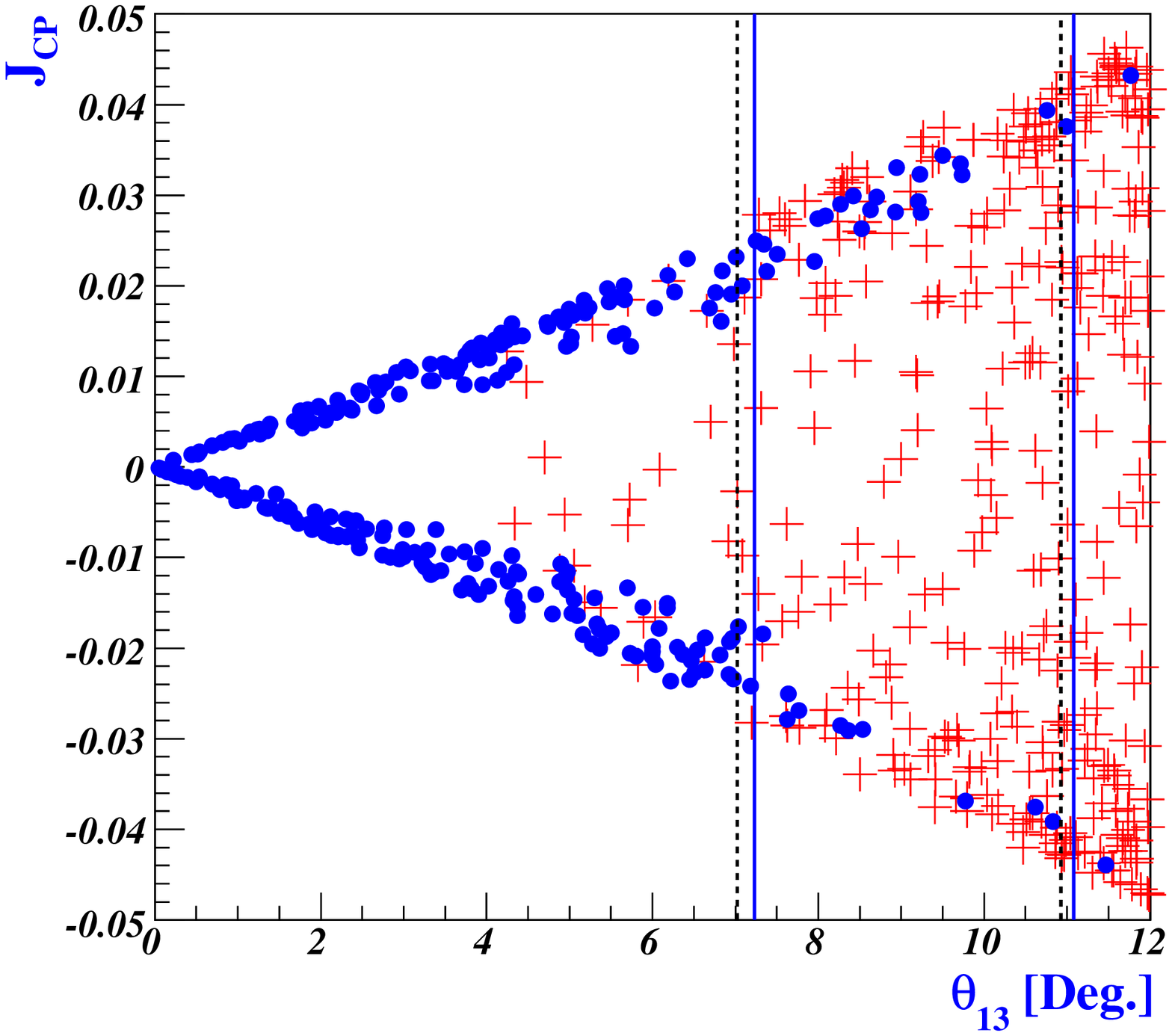,width=6.5cm,angle=0}
\end{minipage}
\hspace*{1.0cm}
\begin{minipage}[t]{6.0cm}
\epsfig{figure=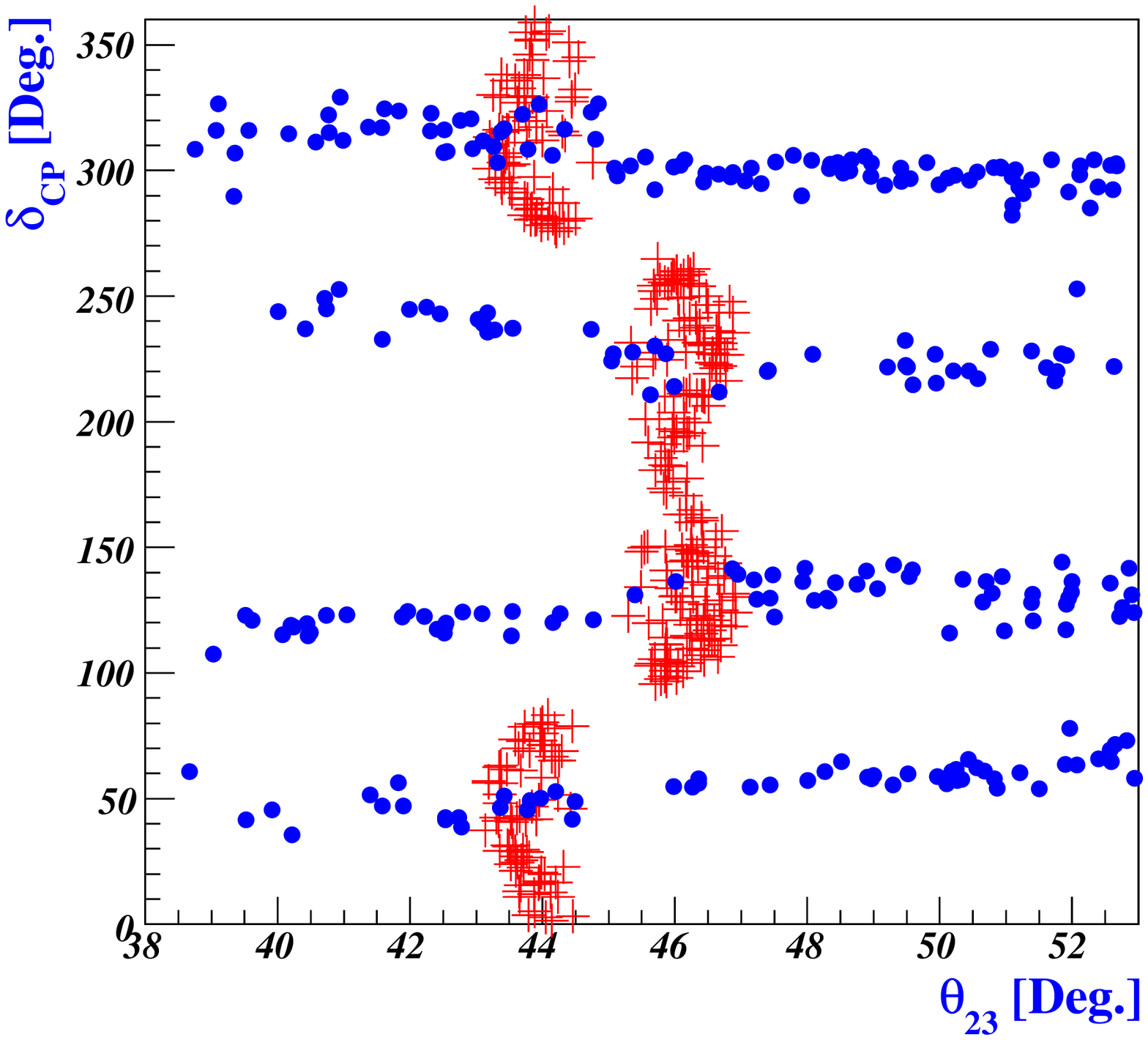,width=6.5cm,angle=0}
\end{minipage}
%\vspace*{-5.5cm}
\caption{\label{FigA1p}
The Jarlskog invariant $J_{CP}$ versus the reactor angle $\theta_{13}$ (left plot), and the Dirac CP phase $\delta_{CP}$ versus $\theta_{23}$ (right plot). The (red) crosses and (blue) dots represent the results for the normal and inverted mass hierarchy, respectively. The vertical solid (dashed) lines in both plots indicate the upper and lower bounds on $\theta_{13}$ for the inverted (normal) mass hierarchy given in Eq.~(\ref{expdata}) at the $3\sigma$ level.}
\end{figure}

\begin{figure}[t]
%\vspace*{-5.0cm}
%\hspace*{-1cm}
\begin{minipage}[t]{6.0cm}
\epsfig{figure=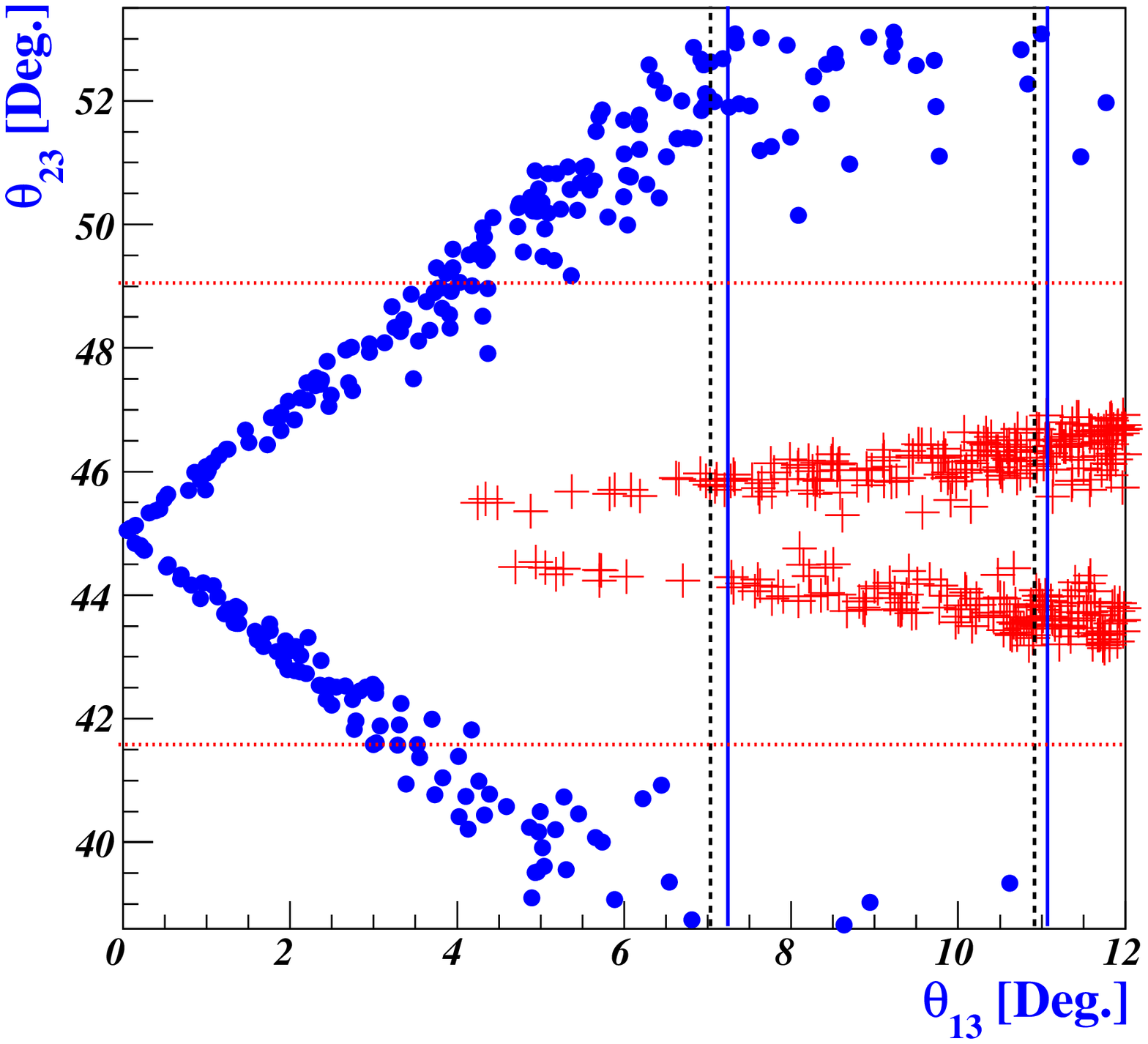,width=6.5cm,angle=0}
\end{minipage}
\hspace*{1.0cm}
\begin{minipage}[t]{6.0cm}
\epsfig{figure=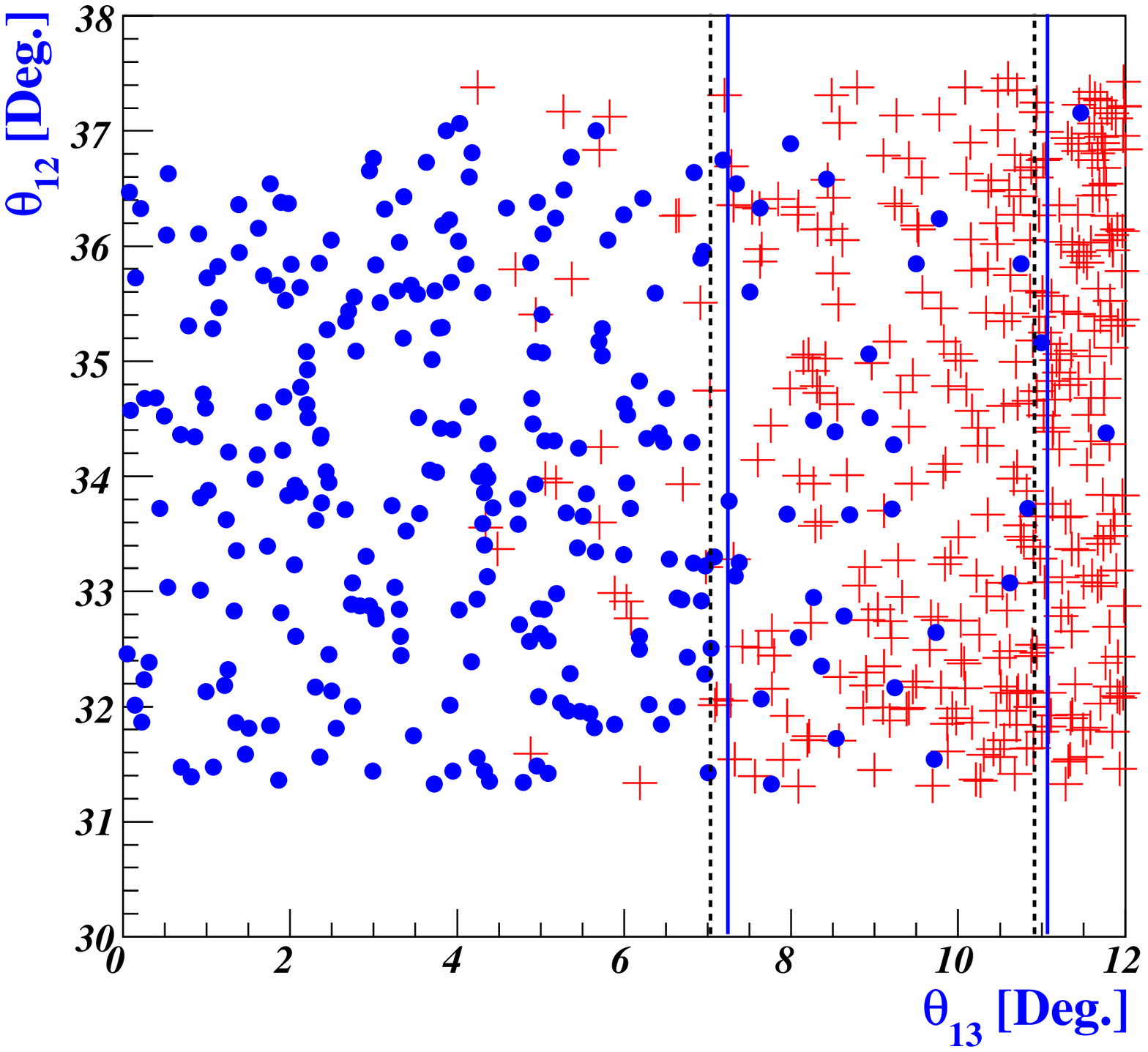,width=6.5cm,angle=0}
\end{minipage}
%\vspace*{-5.5cm}
\caption{\label{FigA2} The behaviors of $\theta_{23}$ and $\theta_{12}$ in terms of  $\theta_{13}$. The red crosses and the blue dots represent results for the normal mass hierarchy and the inverted mass hierarchy, respectively. The solid (dashed) vertical lines represent the experimental bounds of Eq.~(\ref{expnu}) at $3\sigma$'s for the inverted (normal) mass hierarchy. The horizontal dotted lines indicate the $1\sigma$ experimental bounds in Eq.~(\ref{expnu}).
}
\end{figure}

\begin{figure}[t]
%\vspace*{-5.0cm}
%\hspace*{-1cm}
\begin{minipage}[t]{6.0cm}
\epsfig{figure=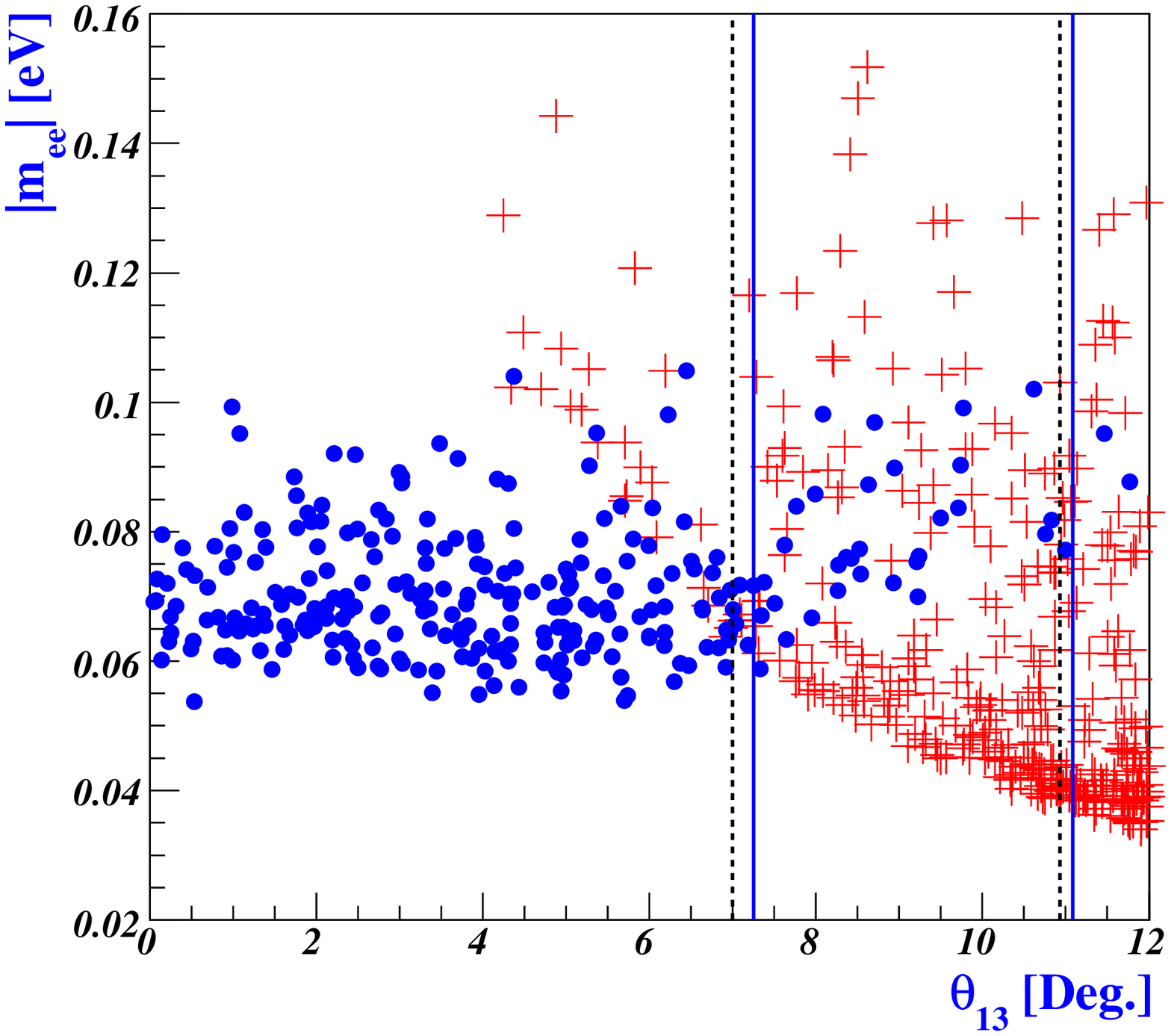,width=6.5cm,angle=0}
\end{minipage}
\hspace*{1.0cm}
\begin{minipage}[t]{6.0cm}
\epsfig{figure=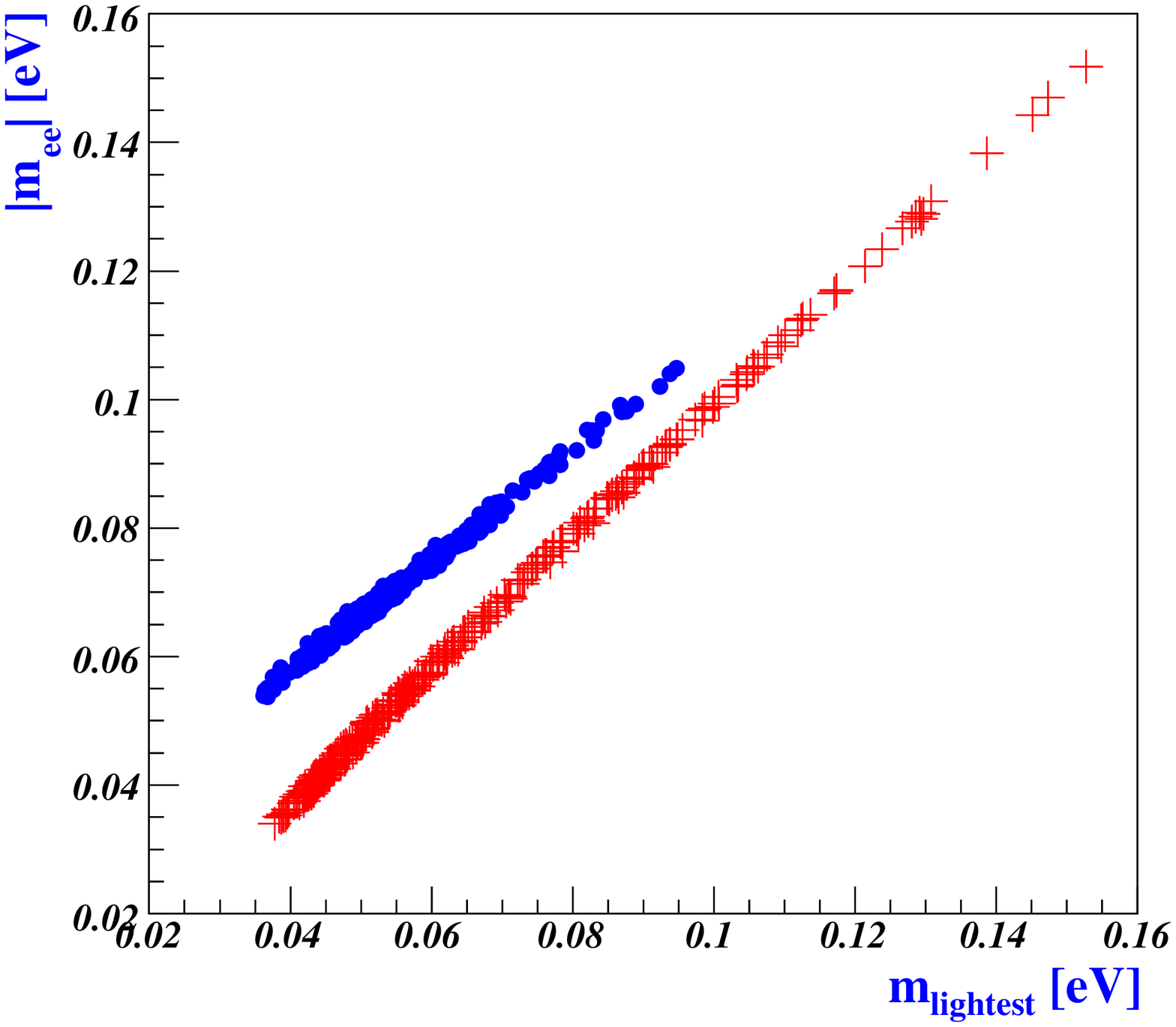,width=6.5cm,angle=0}
\end{minipage}
%\vspace*{-5.5cm}
\caption{\label{FigA3} Plots of $|m_{ee}|$ as a function of $\theta_{13}$ and $m_{\rm lightest}$. The red crosses and the blue dots represent results for the normal and the inverted mass hierarchy, respectively. The vertical solid (dashed) lines show the experimental bounds of Eq.~(\ref{expnu}) at $3\sigma$'s for the inverted (normal) mass hierarchy.}
\end{figure}

In our numerical examples, we take $M=10$ TeV and $v_{\eta}=v_{\Phi}=123$ GeV, for simplicity, as inputs. Since the neutrino masses are sensitive to the combination $m_{0}=v^{2}_{\eta}|y^{\nu2}_{3}|/(6M)$, other choices of $M$ and $v_\eta$ give identical results. Then the parameters $m_{0},y_{1},y_{2},\kappa,\xi$ can be determined from the experimental results of three mixing angles, $\theta_{12},\theta_{13},\theta_{23}$, and the
two mass squared differences, $\Delta m^{2}_{21}, \Delta m^{2}_{31}$. In addition, the CP phases $\delta_{CP},\varphi_{1,2}$ can be predicted after
determining the model parameters.
%%%%%%%%%%%%%%%
%   Fig A-1   %
%%%%%%%%%%%%%%%
\begin{figure}[t]
%\vspace*{-5.0cm}
%\hspace*{-1cm}
\begin{minipage}[t]{6.0cm}
\epsfig{figure=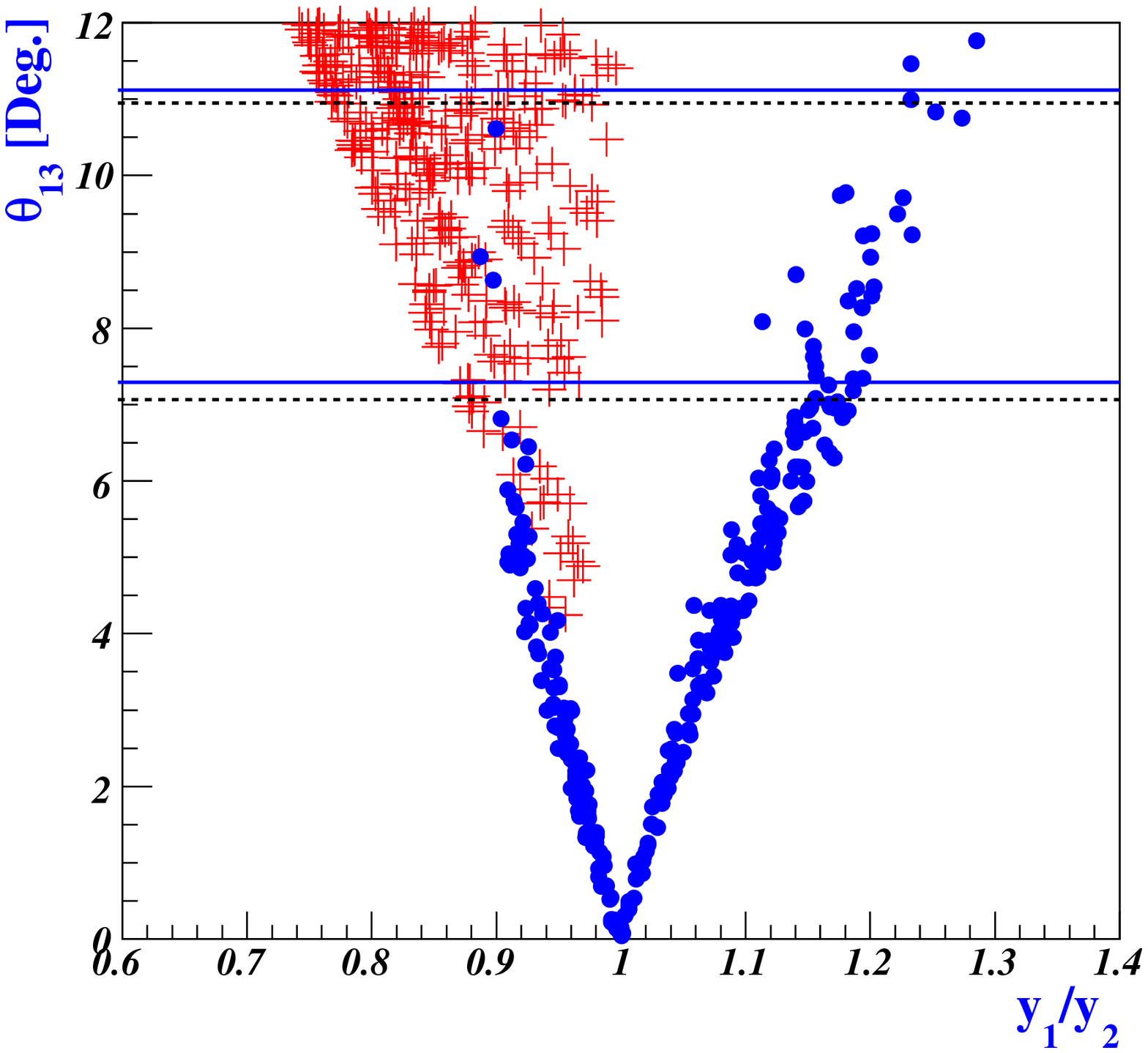,width=6.5cm,angle=0}
\end{minipage}
\hspace*{1.0cm}
\begin{minipage}[t]{6.0cm}
\epsfig{figure=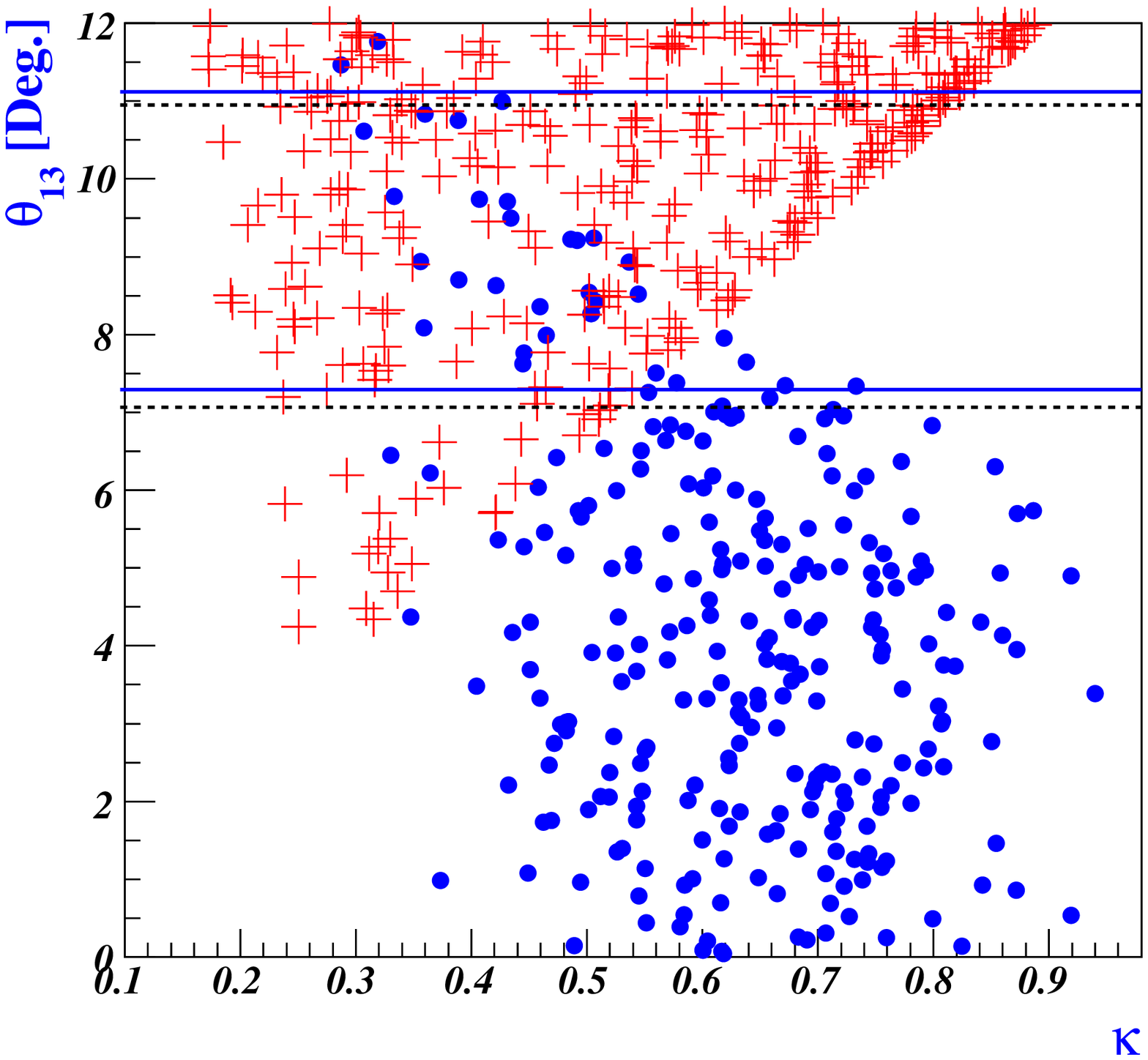,width=6.5cm,angle=0}
\end{minipage}
%\vspace*{-5.5cm}
\caption{\label{FigA1}
The reactor mixing angle $\theta_{13}$ versus the ratio of first-to-second generation neutrino Yukawa couplings $y^{\nu}_{1}/y^{\nu}_{2}$ (left plot) and the parameter $\kappa=|y^{\nu}_{R} v_\chi/M|$ (right plot). The (red) crosses and (blue) dots represent the results for the normal and the inverted mass hierarchy, respectively.  The horizontal solid (dotted) lines in both plots indicate the upper and lower bounds on $\theta_{13}$ for inverted (normal) mass hierarchy given in Eq.~(\ref{expdata}) at the $3\sigma$ level.}
\end{figure}
Using the formulae for the neutrino mixing angles and masses and our values of $M,v_\eta,v_\Phi$, we obtain the following allowed regions of the unknown model parameters: for the normal mass hierarchy (NMH)~\footnote{When $y_{2}=1$ and around there, there exist other parameter spaces giving very small values of $\theta_{13}$. So, we have neglected them in our numerical result for normal mass hierarchy.},
 \begin{eqnarray}
  &&0.17\lesssim\kappa\lesssim0.90~,\qquad0.74\lesssim y_{1}\lesssim1.0~,\qquad0.90\lesssim y_{2}\lesssim1.11~,\nonumber\\
  &&\left\{
      \begin{array}{ll}
        94^{\circ}\lesssim\xi\lesssim119^{\circ}  \\
        240^{\circ}\lesssim\xi\lesssim265^{\circ}
      \end{array}
    \right.
  ~,\qquad\qquad
1.8\lesssim m_{0}\times10^{-2}{\rm [eV]}\lesssim6.0~;
  \label{input1}
 \end{eqnarray}
 for the inverted mass hierarchy (IMH),
 \begin{eqnarray}
  &&0.31\lesssim\kappa\lesssim0.92~,\qquad0.84\lesssim y_{1}\lesssim1.15~,\qquad0.65\lesssim y_{2}\lesssim1.28~,\nonumber\\
  &&\left\{
      \begin{array}{ll}
        90^{\circ}\lesssim\xi\lesssim117^{\circ}  \\
        245^{\circ}\lesssim\xi\lesssim265^{\circ}
      \end{array}
    \right.
  ~,\qquad\qquad1.7\lesssim m_{0}\times10^{-2}{\rm [eV]}\lesssim4.5~.
  \label{input2}
 \end{eqnarray}
Note that here we have used the $3\sigma$ experimental bounds on $\theta_{12},\theta_{23},\Delta m^{2}_{21}, \Delta m^{2}_{31}$ in Eq.~(\ref{expnu}), except for $\theta_{13}<12^{\circ}$ for which we use the values in Eqs.~(\ref{input1},\ref{input2}).
%%%%%%%%%%%%%%%
%   Fig A-1pp   %
%%%%%%%%%%%%%%%
\begin{figure}[t]
%\vspace*{-5.0cm}
%\hspace*{-1cm}
\begin{minipage}[t]{6.0cm}
\epsfig{figure=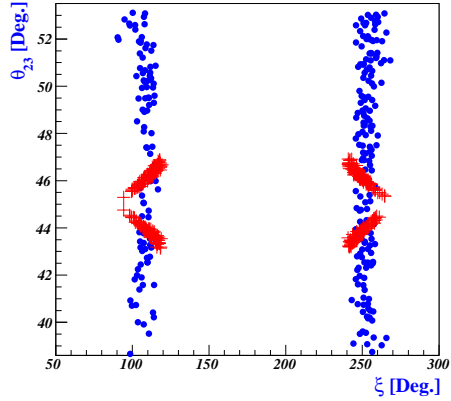,width=6.5cm,angle=0}
\end{minipage}
%\vspace*{-5.5cm}
\caption{\label{FigA1pp}
The atmospheric mixing angle $\theta_{23}$ versus the phase $\xi$ of the parameter combination $y^{\nu}_R v_\chi/M$. The (red) crosses and (blue) dots represent the results for the normal and inverted mass hierarchy, respectively.}
\end{figure}
For these parameter regions, we investigate how a non-zero $\theta_{13}$ can be determined for the normal and inverted mass hierarchy. In Figs.~\ref{FigA1}-\ref{FigA3}, the data points represented by blue dots and red crosses indicate results for the inverted and normal mass hierarchy, respectively. The left-hand-side plot in Fig.~\ref{FigA1} shows how the mixing angle $\theta_{13}$ depends on the ratio $y_1/y_2=y_1^{\nu}/y_2^{\nu}$ of the first- and second-generation neutrino Yukawa couplings; the right-hand-side plot shows how $\theta_{13}$ depends on the parameter $\kappa=|y^{\nu}_R v_\chi/M|$. We see that the measured value of $\theta_{13}$ from the Daya Bay and RENO experiments can be achieved at $3\sigma$'s for $0.75<y_{1}/y_{2}<1$ (NMH), $1.1<y_{1}/y_{2}<1.3$ and $y_{1}/y_{2}\sim0.9$ (IMH), $0.17\lesssim\kappa\lesssim0.82$ (NMH) and $0.3<\kappa\lesssim0.74$ (IMH). Fig.~\ref{FigA1pp} shows the atmospheric mixing angle $\theta_{23}$ as a function of the phase $\xi$ of $y^{\nu}_R v_\chi/M$.
%%%%%%%%%%%%%%%
%   Fig A-1p   %
%%%%%%%%%%%%%%%

To see how the parameters are correlated with low-energy CP violation observables measurable through neutrino oscillations, we consider the leptonic CP violation parameter defined by the
Jarlskog invariant~\cite{Jarlskog:1985ht}
\begin{align}
J_{CP}\equiv{\rm Im}[U_{e1}U_{\mu2}U^{\ast}_{e2}U^{\ast}_{\mu1}]
 =\frac{1}{8}\sin2\theta_{12}\sin2\theta_{23} \sin2\theta_{13}\cos\theta_{13}
 \sin\delta_{CP} .
\end{align}
The Jarlskog invariant $J_{CP}$ can be expressed in terms of the elements of the matrix $h=m_{\nu}m^{\dag}_{\nu}$~\cite{Branco:2002xf}:
 \begin{eqnarray}
  J_{CP}=-\frac{{\rm Im}\{h_{12}h_{23}h_{31}\}}{\Delta m^{2}_{21}\Delta m^{2}_{31}\Delta m^{2}_{32}}~.
  \label{JCP}
 \end{eqnarray}
The behavior of $J_{CP}$ as a function of $\theta_{13}$ is plotted on the left plot of Fig.~\ref{FigA1p}.
We see that the value of $|J_{CP}|$ lies in the range $0-0.04$ (NMH) and $0.02-0.04$ (IMH) for the measured value of $\theta_{13}$ at $3\sigma$'s. Also, in our model we have
 \begin{eqnarray}
  {\rm Im}\{h_{12}h_{23}h_{31}\}=\frac{27m^{6}_{0}}{4a^{4}b^{3}}y^{2}_{1}y^{2}_{2}(1-y^{2}_{2})\sin\psi_{2}\{....\}~,
  \label{JCP1}
 \end{eqnarray}
in which $\{.....\}$ stands for a complicated lengthy function of $y_{1}$, $y_{2}$, $a$, $b$, $\psi_{1}$ and $\psi_{2}$. Clearly, Eq.~(\ref{JCP1}) indicates that in the limit of $y_{2}\rightarrow1$ or $\sin\psi_{2}\rightarrow0$ the leptonic $CP$ violation $J_{CP}$ goes to zero.
 When $y_{2}\neq1$, i.e.\ for the normal hierarchy case, $J_{CP}$ could go to zero as $\sin\psi_{2}$ of Eq.~(\ref{JCP1}).
In the case of the inverted hierarchy, $J_{CP}$ has nonzero values for the measured range of $\theta_{13}$ while $J_{CP}$ goes to zero for $\theta_{13}\rightarrow0$, which corresponds to $y_{2}\rightarrow1$.
The right plot of Fig.~\ref{FigA1p} shows the behavior of the Dirac CP phase $\delta_{CP}$ as a function of $\theta_{23}$, where $\delta_{CP}$ can have discrete values around $50^{\circ},120^{\circ},230^{\circ}$ and $310^{\circ}$ for the inverted mass hierarchy (for the normal mass hierarchy, $\delta_{CP}$ can vary over a wide range except near $90^{\circ}$ and $270^{\circ}$). Future precise measurements of $\theta_{23}$, whether $\theta_{23}>45^{\circ}$ or $\theta_{23}<45^{\circ}$, will provide more information on $\delta_{CP}$.

Fig.~\ref{FigA2} shows how the values of $\theta_{13}$ depend on the mixing angles $\theta_{23}$ and $\theta_{12}$. As can be seen in the left plot of Fig.~\ref{FigA2}, the behavior of $\theta_{23}$ in terms of the measured values of $\theta_{13}$ at $3\sigma$'s for the normal hierarchy is different than for the inverted hierarchy.  For the normal hierarchy we see that the measured values of $\theta_{13}$ can be achieved for $43^{\circ}<\theta_{23}<47^{\circ}$ and $\theta_{23}\neq45^{\circ}$, with small deviations from maximality, while for the inverted hierarchy $50^{\circ}\lesssim\theta_{23}\lesssim53.1^{\circ}$ and $38.6^{\circ}\lesssim\theta_{23}\lesssim40^{\circ}$, which are excluded at $1\sigma$ by the experimental bounds as can be seen in Eq.~(\ref{expnu}).\footnote{Interestingly, the most recent data of MINOS seem to disfavor the maximal mixing in the atmospheric mixing angle, Eq.~(\ref{minos}), indicating that the inverted mass hierarchy may be favored.}
From the right plot of Fig.~\ref{FigA2}, we see that the
predictions for $\theta_{13}$ do not strongly depend on $\theta_{12}$ in the allowed region.
%%%%%%%%%%%%%%%
%   Fig A-2   %
%%%%%%%%%%%%%%%

Moreover, we can straightforwardly obtain the effective neutrino mass $|m_{ee}|$ that characterizes the amplitude for neutrinoless double beta decay :
 \begin{eqnarray}
  |m_{ee}|\equiv \left|\sum_{i}(U_{\rm PMNS})^{2}_{ei}m_{i}\right|~,
  \label{mee}
 \end{eqnarray}
where $U_{\rm PMNS}$ is given in Eq.~(\ref{Unu}).
The left and right plots in Fig.~\ref{FigA3} show the behavior of the effective neutrino mass $|m_{ee}|$ in terms of $\theta_{13}$ and the lightest neutrino mass, respectively.
In the left plot of Fig.~\ref{FigA3}, for the measured values of $\theta_{13}$ at $3\sigma$'s, the effective neutrino mass $|m_{ee}|$ can be in the range $0.04\lesssim|m_{ee}|[{\rm eV}]<0.15$ (NMH) or $0.06\lesssim|m_{ee}|[{\rm eV}]\lesssim0.11$ (IMH).
The right plot of Fig.~\ref{FigA3} shows $|m_{ee}|$ as a function of $m_{\rm lightest}$, where $m_{\rm lightest}=m_{1}$ for the normal mass hierarchy and $m_{\rm lightest}=m_{3}$ for the inverted mass hierarchy.
Our model predicts that the effective mass $|m_{ee}|$ is within the sensitivity of planned neutrinoless double-beta decay experiments.

%%%%%%%%%%%%%%%
%   Fig A-3   %
%%%%%%%%%%%%%%%

%%%%%%%%%%%%%%%%%%%%%%%%%%%%%%%%%%%%%%%%%%%%%%%%%%%%%%%%%%%%%%%%%%%%%%%%%%%%%%%%%%%%
\section{Conclusions}
We have suggested a novel and simple scenario to generate neutrino masses and mixings with a
discrete $A_4$ symmetry that is spontaneously broken. In particular our model can accommodate in a renormalizable Lagrangian a large value of the
mixing angle, $\theta_{13}$, consistent with the recent reactor neutrino experiments Daya Bay and RENO, as well as high energy CP violation interesting for leptogenesis.

In our model we have introduced a right-handed neutrino $N_R$, a real gauge-singlet scalar
$\chi$, and an $SU(2)_L$-doublet scalar $\eta$, all of which are $A_4$ triplets.
The light neutrino masses are generated by a seesaw mechanism in which
we have assumed the right-handed neutrino masses are at the TeV scale (to evade the introduction
of higher dimensional operators). Getting VEVs along the direction $\langle \chi \rangle = v_\chi (1,0,0)$
and $\langle \eta^0 \rangle = v_\eta (1,1,1)$, which break the $A_4$ symmetry
down to a $Z_2$ ($S$-flavor-parity) and a $Z_3$ ($T$-flavor) symmetry, respectively, one obtains bimaximal mixing
at the right-handed neutrino sector and trimaximal mixing at the light Dirac neutrino
sector with non-degenerate Yukawa couplings that deform the exact TBM pattern. The resulting light neutrino mixing matrix is in the form of a deviated TBM generated through unequal neutrino Yukawa couplings, as can be seen
in Figure~\ref{FigA1}. In the limiting case of equal active-neutrino Yukawa couplings, the mixing matrix recovers the exact TBM. In addition, we have shown that unequal neutrino Yukawa couplings can provide a source of high-energy CP violation, perhaps strong enough to be responsible for leptogenesis. The stability of the vacuum alignments we assume are guaranteed, for example, by embedding our model in an extra dimension.

We showed that deviations from the TBM of about 20\% are enough to explain
$\theta_{13} \sim 9^\circ $. We predicted that the CP violating Dirac phase $\delta_{CP}$ may have discrete values (see Figure~\ref{FigA1p}). Therefore the measurement of the phase $\delta_{CP}$ in the
next generation neutrino experiments can rule out or support our model. We have also shown that the inverted mass hierarchy may be excluded by a global analysis using $1\sigma$ experimental bounds, while the most recent MINOS data seem to favor it.
We also predicted an effective neutrino mass in neutrinoless double-beta decay  in the range,
$0.04\lesssim|m_{ee}|[{\rm eV}]<0.15$ (for the normal hierarchy) and $0.06\lesssim|m_{ee}|[{\rm eV}]\lesssim0.11$ (for the inverted hierarchy), both ranges within reach of near-future neutrinoless
double beta decay experiments.

%The residual $T$-flavor symmetry $Z_3$ can make the neutral
%components of $\eta$ stable, and thus good dark matter candidates. A
%detailed analysis of this will be provided elsewhere~\cite{A4DM}.

\newpage
\appendix
%%%%%%%%%%%%%%%%%%%%%%%%%%%%%%%%%%%%%%%%%%%%%%%%%%%%%%%%%%%%%%%%%%%%%%%%%%%%%%%%%%%%%%%%%%%%%%%%%%
\section{The Higgs potential}

\begin{figure}[t]
\vspace*{0.1cm}
 %\hspace*{-2cm}
% \begin{minipage}[t]{6.0cm}
\includegraphics[width=0.40\textwidth]{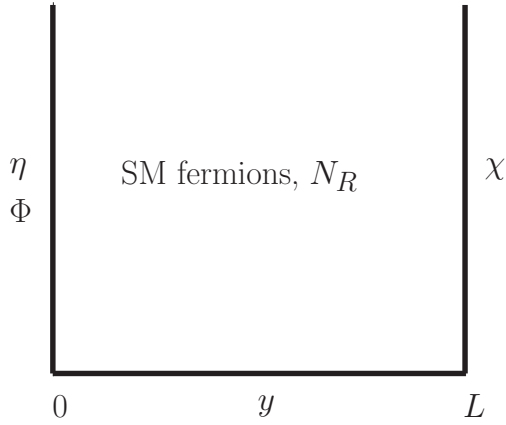}
%  \epsfig{figure=Extra03.eps,width=5cm,height=6cm}
% \end{minipage}
\vspace*{0.1cm}
 \caption{\label{fig:exd}
  Fifth dimension $y$ and locations of scalar and fermion fields on the brane at $y=0$ and $y=L$.}
\end{figure}

In this Appendix we present our Higgs potential and its minimization, as well as our prescription for effecting the stability of the vacuum alignment.
We solve the vacuum alignment problem by extending the model into a spatial extra dimension
$y$~\cite{Altarelli:2005yp}. We assume that each field lives on a 4D brane either at $y = 0$ or at
$y = L$, as shown in Fig.~\ref{fig:exd}. The heavy neutrino masses arise from local operators at $y=0$,
while the charged fermion masses and the neutrino Yukawa interactions are realized by non-local effects
involving both branes
A rigorous explanation of this possibility is beyond the scope of this paper.

The most general renormalizable scalar potential for the Higgs fields $\Phi, \eta$ and $\chi$, invariant under $SU(2)_{L}\times U(1)_{Y}\times A_{4}$ and obeying the conditions in the previous paragraph, is then given by
\begin{eqnarray}
V & =&  V_{y=0}+V_{y=L} ,
\label{potential1}
\end{eqnarray}
where
\begin{eqnarray}
V_{y=0} &=& V(\Phi)+V(\eta)+V(\eta\Phi)~,
\\
V_{y=L}  &=&V(\chi) ~,
\end{eqnarray}
and
 \begin{eqnarray}
V(\eta) &=& \mu^{2}_{\eta}(\eta^{\dag}\eta)_{\mathbf{1}}+\lambda^{\eta}_{1}(\eta^{\dag}\eta)_{\mathbf{1}}(\eta^{\dag}\eta)_{\mathbf{1}}+\lambda^{\eta}_{2}(\eta^{\dag}\eta)_{\mathbf{1^\prime}}(\eta^{\dag}\eta)_{\mathbf{1^{\prime\prime}}}+\lambda^{\eta}_{3}(\eta^{\dag}\eta)_{\mathbf{3}_{s}}(\eta^{\dag}\eta)_{\mathbf{3}_{s}}\\
  &+&\lambda^{\eta}_{4}(\eta^{\dag}\eta)_{\mathbf{3}_{a}}(\eta^{\dag}\eta)_{\mathbf{3}_{a}}+\left\{\lambda^{\eta}_{5}(\eta^{\dag}\eta)_{\mathbf{3}_{s}}(\eta^{\dag}\eta)_{\mathbf{3}_{a}}+\text{h.c.}\right\}~,\nonumber\\
V(\Phi) &=& \mu^{2}_{\Phi}(\Phi^{\dag}\Phi)+\lambda^{\Phi}(\Phi^{\dag}\Phi)^{2}~,\nonumber\\
V(\chi) &=& \mu^{2}_{\chi}(\chi\chi)_{\mathbf{1}}+\lambda^{\chi}_{1}(\chi\chi)_{\mathbf{1}}(\chi\chi)_{\mathbf{1}}+\lambda^{\chi}_{2}(\chi\chi)_{\mathbf{1}^\prime}(\chi\chi)_{\mathbf{1}^{\prime\prime}}+\lambda^{\chi}_{3}(\chi\chi)_{\mathbf{3}_{s}}(\chi\chi)_{\mathbf{3}_{s}}\\
  &+&\lambda^{\chi}_{4}(\chi\chi)_{\mathbf{3}_{a}}(\chi\chi)_{\mathbf{3}_{a}}
  +\lambda^{\chi}_{5}(\chi\chi)_{\mathbf{3}_{s}}(\chi\chi)_{\mathbf{3}_{a}}+\xi^{\chi}_{1}\chi(\chi\chi)_{\mathbf{3}_{s}}+\xi^{\chi}_{2}\chi(\chi\chi)_{\mathbf{3}_{a}}~,\nonumber\\
V(\eta\Phi) &=& \lambda^{\eta\Phi}_{1}(\eta^{\dag}\eta)_{\mathbf{1}}(\Phi^{\dag}\Phi)
  +\lambda^{\eta\Phi}_{2}[(\eta^{\dag}\Phi)(\Phi^{\dag}\eta)]_{\mathbf{1}}+\left\{\lambda^{\eta\Phi}_{3}[(\eta^{\dag}\Phi)(\eta^{\dag}\Phi)]_{\mathbf{1}}+\text{h.c.}\right\}\\
  &+&\left\{\lambda^{\eta\Phi}_{4}(\eta^{\dag}\eta)_{\mathbf{3}_{s}}(\eta^{\dag}\Phi)+\text{h.c.}\right\}+\left\{\lambda^{\eta\Phi}_{5}(\eta^{\dag}\eta)_{\mathbf{3}_{a}}(\eta^{\dag}\Phi)+\text{h.c.}\right\}~.
\label{potential6}
\end{eqnarray}
Here $\mu_{\eta},\mu_{\Phi},\mu_{\chi}$, $\xi^{\chi}_{1}$ and $\xi^{\chi}_{2}$ have mass dimension-1, while $\lambda^{\eta}_{1,...,5}$, $\lambda^{\Phi}$, $\lambda^{\chi}_{1,...,5}$ and $\lambda^{\eta\Phi}_{1,...,5}$ are  dimensionless. In $V(\eta\Phi)$ the usual mixing term $\Phi^{\dag}\eta$ is forbidden by the $A_{4}$ symmetry. In the scalar potential (\ref{potential1})--(\ref{potential6})  we have for simplicity assumed that $CP$ is conserved, and the couplings $\lambda^{\eta\Phi}_{3},\lambda^{\eta\Phi}_{4}$ and $\lambda^{\eta}_{5}$ are real.

%%%%%%%%%%%%%%%
%   Fig Appendix   %
%%%%%%%%%%%%%%%
%\vspace{0.5cm}
%%%%%%%%%%%%%%%%%%%%%%%%%%%%%%%%%%%%%%%

The vacuum configuration is obtained by the vanishing of the derivative of $V$ with respect to each component of the scalar fields $\Phi,\eta_{i}, \chi_{i}~(i=1,2,3)$. The vacuum alignment of the field $\eta$ is determined by
 \begin{eqnarray}
\sqrt{2}\, \frac{\partial V}{\partial \eta^{0}_{1}}\Big|_{\langle\eta^{0}_{i}\rangle=v_{\eta_{i}}}&=& \frac{v_{\eta_{1}}}{2}\Big\{2v^{2}_{\eta_{1}}(\lambda^{\eta}_{1}+\lambda^{\eta}_{2})+2\mu^{2}_{\eta}+(v^{2}_{\eta_{2}}+v^{2}_{\eta_{3}})(2\lambda^{\eta}_{1}-\lambda^{\eta}_{2}+4\lambda^{\eta}_{3})\nonumber\\
  &+&v^{2}_{\Phi}(\lambda^{\eta\Phi}_{1}+\lambda^{\eta\Phi}_{2}+2\lambda^{\eta\Phi}_{3})\Big\}+3v_{\eta_{2}}v_{\eta_{3}}v_{\Phi}\lambda^{\eta\Phi}_{4}=0~,\nonumber\\
\sqrt{2}\,  \frac{\partial V}{\partial \eta^{0}_{2}}\Big|_{\langle\eta^{0}_{i}\rangle=v_{\eta_{i}}}&=& \frac{v_{\eta_{2}}}{2}\Big\{2v^{2}_{\eta_{2}}(\lambda^{\eta}_{1}+\lambda^{\eta}_{2})+2\mu^{2}_{\eta}+(v^{2}_{\eta_{1}}+v^{2}_{\eta_{3}})(2\lambda^{\eta}_{1}-\lambda^{\eta}_{2}+4\lambda^{\eta}_{3})\nonumber\\
  &+&v^{2}_{\Phi}(\lambda^{\eta\Phi}_{1}+\lambda^{\eta\Phi}_{2}+2\lambda^{\eta\Phi}_{3})\Big\}+3v_{\eta_{1}}v_{\eta_{3}}v_{\Phi}\lambda^{\eta\Phi}_{4}=0~,\nonumber\\
\sqrt{2}\, \frac{\partial V}{\partial \eta^{0}_{3}}\Big|_{\langle\eta^{0}_{i}\rangle=v_{\eta_{i}}}&=& \frac{v_{\eta_{3}}}{2}\Big\{2v^{2}_{\eta_{3}}(\lambda^{\eta}_{1}+\lambda^{\eta}_{2})+2\mu^{2}_{\eta}+(v^{2}_{\eta_{1}}+v^{2}_{\eta_{2}})(2\lambda^{\eta}_{1}-\lambda^{\eta}_{2}+4\lambda^{\eta}_{3})\nonumber\\
  &+&v^{2}_{\Phi}(\lambda^{\eta\Phi}_{1}+\lambda^{\eta\Phi}_{2}+2\lambda^{\eta\Phi}_{3})\Big\}+3v_{\eta_{1}}v_{\eta_{2}}v_{\Phi}\lambda^{\eta\Phi}_{4}=0~.
 \end{eqnarray}
From this set of three equations, we obtain the solution
 \begin{eqnarray}
  \langle\eta^{0}_{1}\rangle&=&\langle\eta^{0}_{2}\rangle=\langle\eta^{0}_{3}\rangle\equiv v_{\eta}\nonumber\\
  &=&\frac{-3v_{\Phi}\lambda^{\eta\Phi}_{4}\pm\sqrt{9v^{2}_{\Phi}\lambda^{\eta\Phi2}_{4}-2(3\lambda^{\eta}_{1}+4\lambda^{\eta}_{3})(2\mu^{2}_{\eta}+v^{2}_{\Phi}(\lambda^{\eta\Phi}_{1}+\lambda^{\eta\Phi}_{2}+2\lambda^{\eta\Phi}_{3}))}}{2(3\lambda^{\eta}_{1}+4\lambda^{\eta}_{3})}\neq0~.
  \label{veveta}
 \end{eqnarray}
This VEV breaks $A_{4}$ down to a residual $Z_{3}$.

The vanishing of the derivative of $V$ with respect to $\Phi$ reads
 \begin{eqnarray}
\sqrt{2}\, \frac{\partial V}{\partial \Phi^{0}}\Big|_{\langle\Phi^{0}\rangle=v_{\Phi}}&=& v_{\Phi}\left\{v^{2}_{\Phi}\lambda^{\Phi}+\mu^{2}_{\Phi}+\frac{1}{2}(\lambda^{\eta\Phi}_{1}+\lambda^{\eta\Phi}_{2}+2\lambda^{\eta\Phi}_{3})(v^{2}_{\eta1}+v^{2}_{\eta2}+v^{2}_{\eta3})\right\}\nonumber\\
  &+&3v_{\eta_{1}}v_{\eta_{2}}v_{\eta_{3}}\lambda^{\eta\Phi}_{4}=0~.
  \label{dvdphi0}
 \end{eqnarray}
 %\pg{(I removed a $v_\Phi$ in the last term.)}
The real-valued solution of Eq.~(\ref{dvdphi0}), for real-valued parameters, is
 \begin{eqnarray}
  v_{\Phi}=\frac{-(\frac{2}{3})^{1/3}\tilde{b}}{\left\{-9\tilde{a}^{2}\tilde{c}+\sqrt{3(4\tilde{a}^{3}\tilde{b}^{3}+27\tilde{a}^{4}\tilde{c}^{2})}\right\}^{1/3}}+\frac{\left\{-9\tilde{a}^{2}\tilde{c}+\sqrt{3(4\tilde{a}^{3}\tilde{b}^{3}+27\tilde{a}^{4}\tilde{c}^{2})}\right\}^{1/3}}{\tilde{a}(18)^{1/3}}~,
 \end{eqnarray}
where $\tilde{a}=\lambda^{\Phi}, \tilde{b}=\mu^{2}_{\Phi}+\frac{3}{2}v^{2}_{\eta}(\lambda^{\eta\Phi}_{1}+\lambda^{\eta\Phi}_{2}+2\lambda^{\eta\Phi}_{3})$ and $\tilde{c}=3v^{3}_{\eta}\lambda^{\eta\Phi}_{4}$.

Finally, the minimization equations for the vacuum configuration of $\chi$ are given by
 \begin{eqnarray}
  \frac{\partial V}{\partial \chi_{1}}\Big|_{\langle\chi_{i}\rangle=v_{\chi_{i}}}&=&  2v_{\chi_{1}}\Big(\mu^{2}_{\chi}+(2\lambda^{\chi}_{1}-\lambda^{\chi}_{2}+4\lambda^{\chi}_{3})(v^{2}_{\chi_{2}}+v^{2}_{\chi_{3}})
  +2(\lambda^{\chi}_{1}+\lambda^{\chi}_{2})v^{2}_{\chi_{1}}\Big)\nonumber\\
  &+&6\xi^{\chi}_{1}v_{\chi_{2}}v_{\chi_{3}}=0~,\nonumber\\
  \frac{\partial V}{\partial \chi_{2}}\Big|_{\langle\chi_{i}\rangle=v_{\chi_{i}}}&=&  2v_{\chi_{2}}\Big(\mu^{2}_{\chi}+(2\lambda^{\chi}_{1}-\lambda^{\chi}_{2}+4\lambda^{\chi}_{3})(v^{2}_{\chi_{1}}+v^{2}_{\chi_{3}})
  +2(\lambda^{\chi}_{1}+\lambda^{\chi}_{2})v^{2}_{\chi_{2}}\Big)\nonumber\\
  &+&6\xi^{\chi}_{1}v_{\chi_{1}}v_{\chi_{3}}=0~,\nonumber\\
  \frac{\partial V}{\partial \chi_{3}}\Big|_{\langle\chi_{i}\rangle=v_{\chi_{i}}}&=&  2v_{\chi_{3}}\Big(\mu^{2}_{\chi}+(2\lambda^{\chi}_{1}-\lambda^{\chi}_{2}+4\lambda^{\chi}_{3})(v^{2}_{\chi_{1}}+v^{2}_{\chi_{2}})
  +2(\lambda^{\chi}_{1}+\lambda^{\chi}_{2})v^{2}_{\chi_{3}}\Big)\nonumber\\
  &+&6\xi^{\chi}_{1}v_{\chi_{1}}v_{\chi_{2}}=0~.
 \end{eqnarray}
From these equations, we obtain the solution\footnote{There exists another nontrivial solution $\langle\chi\rangle=v_{\chi}(1,1,1)$ with $v_{\chi}=\frac{-3\xi^{\chi}_{1}\pm\sqrt{9\xi^{\chi2}_{1}-8\mu^{2}_{\chi}(3\lambda^{\chi}_{1}+4\lambda^{\chi}_{3})}}{4(3\lambda^{\chi}_{1}+4\lambda^{\chi}_{3})}$.
But this solution is not of interest for our purposes.}
 \begin{eqnarray}
  \langle\chi_{1}\rangle\equiv v_{\chi}=\sqrt{\frac{-\mu^{2}_{\chi}}{2(\lambda^{\chi}_{1}+\lambda^{\chi}_{2})}}\neq0~,\quad\langle\chi_{2}\rangle=\langle\chi_{3}\rangle=0~.
  \label{vevchi}
 \end{eqnarray}

%%%%%%%%%%%%%%%%%%%%%%%%%%%%%%%%%%%%%%%%%%%%%%%%%%%%%%%%%%%%%%%%%%%%%%%%%%%%%%%%%%%%%%%%%%%%%%%%%%%
%\subsection{Comments on section II}
%\label{trans}
%After the breaking of the flavor and electroweak symmetries, the residual $Z_{2}$ symmetry in the heavy Majorana and Higgs sectors is defined under $S$ as
% \begin{eqnarray}
%&&N_{R1} \to N_{R1}, \qquad\qquad~~ \chi_{1} \to \chi_{1}\nonumber\\
%&&N_{R2} \to -N_{R2}, \qquad\qquad \chi_{2} \to -\chi_{2}\nonumber\\
%&&N_{R3} \to -N_{R3}, \qquad\qquad \chi_{3} \to -\chi_{3}~.
% \end{eqnarray}
%
%And the $Z_{3}$ residual symmetry in neutrino Yukawa and Higgs sectors is defined under $T$ as
% \begin{eqnarray}
% &&N_{R1}\rightarrow N_{R2}~,\qquad h_{1}~(A_{1})\rightarrow h_{2}~(A_{2})~,\quad\quad~ \nu_{e}~(e)\rightarrow\nu_{e}~(e)~,\nonumber\\
% &&N_{R2}\rightarrow N_{R3}~,\qquad h_{2}~(A_{2})\rightarrow h_{3}~(A_{3})~,\qquad \nu_{\mu}~(\mu)\rightarrow\omega\nu_{\mu}~(\omega\mu)~,\nonumber\\
%  &&N_{R3}\rightarrow N_{R1}~,\qquad h_{3}~(A_{3})\rightarrow  h_{1}~( A_{1})~,\qquad \nu_{\tau}~(\tau)\rightarrow\omega^{2}\nu_{\tau}~(\omega^{2}\tau)~.
% \label{int}
% \end{eqnarray}

%%%%%%%%%%%%%%%%%%%%%%%%%%%%%%%%%%%%%%%%%%%%%%%%%%%%%%%%%%%%%%%%%%%%%%%%%%%%%%%%%%%%%%%%%%%%%%%%%%
\section{Parametrization of the neutrino mass matrix}

We parametrize the hermitian matrix $m_{\nu}m^{\dag}_{\nu}$ as follows:
 \begin{eqnarray}
 m_{\nu}m^{\dag}_{\nu}= m^{2}_{0}\left(\begin{array}{ccc}
  \tilde{A}y^{2}_{1} & y_{1}y_{2}\left(\frac{3Q-P}{2}-i\frac{3(R-S)}{2}\right) & y_{1}\left(-\frac{3Q+P}{2}-i\frac{3(R+S)}{2}\right) \\
  y_{1}y_{2}\left(\frac{3Q-P}{2}+i\frac{3(R-S)}{2}\right) & y^{2}_{2}\left(F+G+\frac{9K}{4}+\frac{3D}{2}\right) & y_{2}\left(F+G-\frac{9K}{4}-i\frac{3Z}{2}\right) \\
  y_{1}\left(-\frac{3Q+P}{2}+i\frac{3(R+S)}{2}\right) & y_{2}\left(F+G-\frac{9K}{4}+i\frac{3Z}{2}\right) & F+G+\frac{9K}{4}+\frac{3D}{2}
  \end{array}\right).~\nonumber
 \end{eqnarray}
 All parameters appearing here are real, and equal to
 \begin{eqnarray}
  \tilde{A}&=&1+y^{2}_{1}+y^{2}_{2}+\frac{1+4y^{2}_{1}+y^{2}_{2}}{a^{2}}-\frac{2(1-2y^{2}_{1}+y^{2}_{2})\cos\psi_{1}}{a}~,\qquad K=\frac{1+y^{2}_{2}}{b^{2}}~,\nonumber\\
  F&=& 1+y^{2}_{1}+y^{2}_{2}+\frac{1+4y^{2}_{1}+y^{2}_{2}}{4a^{2}}~,\qquad \qquad G=\frac{(1-2y^{2}_{1}+y^{2}_{2})\cos\psi_{1}}{a}~,\nonumber\\
  D&=& (1-y^{2}_{2})\frac{\cos\psi_{12}+2a\cos\psi_{2}}{ab}~,\qquad\qquad Z=(1-y^{2}_{2})\frac{\sin\psi_{12}-2a\sin\psi_{2}}{ab}~,\nonumber\\
  P&=& \frac{1+4y^{2}_{1}+y^{2}_{2}}{a^{2}}-2(1+y^{2}_{1}+y^{2}_{2})+\frac{(1-2y^{2}_{1}+y^{2}_{2})\cos\psi_{1}}{a}~,\nonumber\\
  Q&=& (1-y^{2}_{2})\frac{\cos\psi_{12}-a\cos\psi_{2}}{ab}~,\qquad~\qquad S=(1-y^{2}_{2})\frac{\sin\psi_{12}+a\sin\psi_{2}}{ab}~,\nonumber\\
  R&=&\frac{(1-2y^{2}_{1}+y^{2}_{2})\sin\psi_{1}}{a}~.
 \label{MMele}
 \end{eqnarray}

In Eq.~(\ref{sol1}) the parameters $\Psi_{1},\Psi_{2}$ are defined by
 \begin{eqnarray}
  \Psi_{1}&=&c^{2}_{13}y^{2}_{1}\tilde{A}+s^{2}_{13}\Big\{\left(F+G+\frac{9K}{4}+\frac{3D}{2}\right)(c^{2}_{23}+y^{2}_{2}s^{2}_{23})+y_{2}\left(F+G-\frac{9K}{4}\right)\sin2\theta_{23}\Big\}~\nonumber\\
  &-&\frac{y_{1}}{2}\sin2\theta_{13}\Big\{3c_{23}(R+S)\sin\delta_{CP}-(3Q+P)\cos\delta_{CP}\nonumber\\
  &+&y_{2}s_{23}\left(3(R-S)\sin\delta_{CP}+(3Q-P)\cos\delta_{CP}\right)\Big\}\nonumber\\
  \Psi_{2}&=&\left(F+G+\frac{9K}{4}+\frac{3D}{2}\right)(s^{2}_{23}+y^{2}_{2}c^{2}_{23})-y_{2}\left(F+G-\frac{9K}{4}\right)\sin2\theta_{23}~.
 \label{para1}
 \end{eqnarray}

%%%%%%%%%%%%%%%%%%%%%%%%%%%%%%%%%%%%%%%%%%%%%%%%%%%%%%%%%%%%%%%%%%%%%%%%%%%%%%%%%%%%
\acknowledgments{
This work was supported by NRF Research Grant  2012R1A2A1A01006053 (SB). PG was supported in part by NSF grant PHY-1068111 at the University of Utah and thanks the Korean Instituted for Advanced Studies and Seoul National University for support during the completion of this work.
}

%%%%%%%%%%%%%%%%%%%%%%%%%%%%%%%%%%%%%%%%%%%%%%%%%%%%%%%%%%%%%%%%%%%%%%%%%%%%%%%%%%%%%

\end{document}